\documentclass[smallextended]{svjour3}

\usepackage[dvipsnames]{xcolor}
\usepackage{pifont}
\usepackage{amsmath,amssymb,amsfonts}
\usepackage{algorithmic}
\usepackage{graphicx}
\usepackage{url}
\usepackage{textcomp}
\usepackage{fixltx2e}
\def\BibTeX{{\rm B\kern-.05em{\sc i\kern-.025em b}\kern-.08em
    T\kern-.1667em\lower.7ex\hbox{E}\kern-.125emX}}

\usepackage{listings}
\usepackage{xspace}
\usepackage{subcaption}

\usepackage{color,soul}
\usepackage{adjustbox}
\usepackage{gensymb}
\usepackage{booktabs} 

\usepackage[htt]{hyphenat}

\usepackage{wrapfig}
\usepackage{tikz}
\usetikzlibrary{decorations}
\usetikzlibrary{backgrounds}
\usetikzlibrary{arrows,shapes}
\usetikzlibrary{tikzmark}
\usetikzlibrary{calc}
\newcommand{\highlight}[2]{\colorbox{#1!17}{$\displaystyle #2$}}

\usepackage{tcolorbox}
\definecolor{dkgreen}{rgb}{0,0.6,0}
\definecolor{gray}{rgb}{0.5,0.5,0.5}
\definecolor{mauve}{rgb}{0.58,0,0.82}

\newcommand{\threeStars}{\mbox{$^{***}$}}
\newcommand{\twoStars}{\mbox{$^{** }$}}
\newcommand{\oneStar}{\mbox{$^{*  }$}}

\usepackage{booktabs} 
\usepackage{amsmath}

\begin{document}

\title{Opportunities and Security Risks of Technical Leverage: A Replication Study on the NPM Ecosystem}





\author{Haya Samaana, Diego Elias Costa, Ahmad Abdellatif, and Emad Shihab}

\institute{ 
Haya Samaana\\
Department of Computer Engineering 
\at An-Najah National University, Nablus, Palestine\\
\email{hayasam@najah.edu}\\
Diego Elias Costa\\
Department of Computer Science and Software Engineering
\at Concordia University, Montreal, Canada\\
\email{diego.costa@concordia.ca}\\
Ahmad Abdellatif\\ 
Department of Electrical \& Software Engineering
\at University of Calgary, Calgary, Canada \\
\email{ahmad.abdellatif@ucalgary.ca}\\
Emad Shihab \\
Department of Computer Science and Software Engineering
\at Concordia University, Montreal, Canada\\
\email{emad.shihab@concordia.ca}
}

\maketitle
\begin{abstract}

To comply with high productivity demands, software developers reuse free open-source software (FOSS) code to avoid reinventing the wheel when incorporating software features. The reliance on FOSS reuse has been shown to improve productivity and the quality of delivered software; however, reusing FOSS comes at the risk of exposing software projects to public vulnerabilities. Massacci and Pashchenko have explored this trade-off in the Java ecosystem through the lens of technical leverage: the ratio of code borrowed from FOSS over the code developed by project maintainers. 

In this paper, we replicate the work of  Massacci and Pashchenko and we expand the analysis to include level-1 transitive dependencies to study technical leverage in the fastest-growing NPM ecosystem. We investigated 14,042 NPM library releases and found that both opportunities and risks of technical leverage are magnified in the NPM ecosystem. Small-medium libraries leverage 2.5x more code from FOSS than their code, while large libraries leverage only 3\% of FOSS code in their projects. Our models indicate that technical leverage shortens the release cycle for small-medium libraries. However, the risk of vulnerability exposure is 4-7x higher for libraries with high technical leverage. 
We also expanded our replication study to include the first level of transitive dependencies, and show that the results still hold, albeit with significant changes in the magnitude of both opportunities and risks of technical leverage.  

Our results indicate the extremes of opportunities and risks in NPM, where high technical leverage enables fast releases but comes at the cost of security risks.

\end{abstract}


\section{Introduction} \label{intro}


Software ecosystems have transformed how we develop software. They enable developers to publish their code as software libraries that can be easily leveraged by other software developers.
For example, the NPM ecosystem alone has more than 2.15M libraries that JavaScript developers can use.  
The numerous free open-source software libraries facilitate and encourage code reusability. 
By leveraging those libraries, developers avoid reinventing the wheel~\cite{grinter1996understanding}, improve software quality~\cite{mohagheghi2004empirical,basili1996reuse,abdellatif_ist2020} and reduce time-to-market~\cite{mohagheghi2007quality}. 


While leveraging FOSS libraries brings numerous opportunities for speeding up software development, the risk of borrowing other developers' code manifests in the form of bugs and vulnerabilities. 
Prior work shows that vulnerabilities are a widespread problem in FOSS libraries \cite{dashevskyi2018screening,Alfadel:2021:Python,Alfadel21Dependabot}.
Numerous security incidents point to vulnerable dependencies as the main culprit for the security exploit. 
For example, an exploit in the vulnerability found in Apache Structs led to a data breach of Equifax systems, leaking data of millions of American citizens and leading to a cost of 1.8 billion US dollars in settlements and security upgrades~\cite{luszcz2018apache,Equifaxd60:online}.  
More recently, a vulnerability in the log4j library affected a plethora of services and systems, including e-commerce websites overall the world~\cite{neuburger2021trends,yu2022pbdiff}.


To develop software efficiently, developers have to constantly deal with the trade-off of the opportunities brought by reusing FOSS libraries versus the security risks of depending on other people's code.
Recently, Massacci and Pashchenko~\cite{massacci2021technical} have modelled this trade-off using the notion of \textbf{technical leverage}. 
Technical leverage expresses the ratio of a software project code that is borrowed from FOSS libraries over the size of one's own code. 
The authors investigate the opportunities and risks of technical leverage in the Maven ecosystem (Java), reporting a clear trade-off:  
Leveraging FOSS code (high technical leverage) can help projects ship more code without incurring large release delays, however, projects with such high technical leverage are also 60\% more likely to become vulnerable than projects with low technical leverage. 


The work of Massacci and Pashchenko~\cite{massacci2021technical} has inspired us to replicate their analysis on another major software ecosystem - the NPM ecosystem - for the following reasons: 
\textbf{1)} NPM is the package manager for Node JS libraries in JavaScript, the most popular programming language among developers\footnote{https://insights.stackoverflow.com/survey/2021}, and is the largest and fastest growing ecosystem to date. NPM has more than 2 million reusable packages~\cite{Libraries,latendresse2022not} and has very distinct characteristics to the Maven ecosystem.
\textbf{2)} Due to the minimalist JavaScript standard library, JavaScript developers tend to reuse more FOSS components in their projects (high opportunity for technical leverage)~\cite{decan2018evolution,zerouali2018empirical}. 
\textbf{3)} 
Vulnerabilities are very commonly found in NPM packages. 
Studies and industry reports estimate that between 30\% to 40\% of all NPM packages rely on code with known vulnerabilities (high risk of technical leverage)~\cite{Joseph_Hejderup,2020Soft27:online}.
\textbf{4)} Studies have frequently pointed to major differences across software ecosystems regarding software reuse~\cite{Alfadel:2021:Python,decan2017empirical,Bogart:21:PracticesPolicies}.
Therefore, what are the differences between the opportunities and risks of technical leverage between Maven and the NPM ecosystem?

In this study, we assess the opportunities and risks of technical leverage on popular packages of the NPM ecosystem. 
First, we focus on studying the direct technical leverage for 14,042 releases, by considering only the direct dependencies of libraries, thus disregarding transitive dependencies to compare our results with Massacci and Pashchenko~\cite{massacci2021technical}.
Then, we expand on the replicated study to include an assessment of the opportunities and risks of transitive dependencies, not initially included in the replicated article.
To keep the computational costs manageable, we included the first level of transitive dependencies and report on the impact of its inclusion on all analysis performed by Massacci and Pashchenko~\cite{massacci2021technical}.

Our study aims to answer the following research questions: 

\textbf{RQ1: Is there a difference in direct technical leverage, distance and direction of changes between small-medium and large libraries?}
We analyze how small-medium libraries($<$10KLOC) and large libraries\\($\geq$10KLOC) leverage FOSS and their related maintenance activities over time.
Our findings show that small-medium libraries tend to leverage multiple times their code in FOSS (2.5x in median), while large libraries leverage only 3\% of their code base in FOSS. 
When we look at the evolution of these libraries, results show that small-medium libraries constantly adopt new dependencies across releases, while developers of large libraries mostly work on their own code.  
While similar results were found in the Maven ecosystem~\cite{massacci2021technical}, we find that, contrary to our beliefs, NPM libraries have lower direct technical leverage than Maven libraries.




\textbf{RQ2: How does direct technical leverage impact the time interval between library releases?}
We build a multivariate linear regression model to capture how direct technical leverage is associated with the time interval between library releases. 
Our results confirm our intuition:
High direct technical leverage has a positive association with faster release cycles for small-medium libraries. 
The positive effect was considerably higher in NPM libraries than the effect reported for Maven libraries in the replicated study~\cite{massacci2021technical}. 
However, technical leverage has no significant effect on large libraries' release cycles.


\textbf{RQ3: Does direct technical leverage impact the risk of including more vulnerabilities?}
We explore the relation between direct technical leverage and risks in terms of vulnerabilities. Therefore, we use the Odds Ratio (OR) to quantify the risk of high direct technical leverage on libraries. 
We find that high direct technical leverage brings high risks. 
In particular, small-medium libraries have 4 times more chances of being vulnerable if they have high technical leverage. 
In large libraries, high technical leverage increases the risk of vulnerability to 7 folds. In the Maven ecosystem, small-medium libraries have 1.6 higher odds of being
vulnerable in comparison to the large libraries with OR = 0.43 \cite{massacci2021technical}.


\textbf{RQ4: To what extent do the findings about the technical leverage observed for direct dependencies hold for level-1 transitive dependencies?}
We complement our study by investigating the first level of transitive dependencies to understand its impact on the difference in technical leverage between small-medium
and large libraries, the time interval between library releases, and the risk of including more vulnerabilities.\\
Results show similar opportunities and risks, compared to considering only direct dependencies, but with a significant change in the magnitude of both benefits and downsides. 
Expectedly, adding the first level of transitive dependencies in the analysis increases significantly the technical leverage. In particular, the technical leverage of a small-medium library rose from 2.5 to 7.7 times their own code in FOSS.  
As a consequence of this increase in technical leverage, the risks of vulnerabilities also saw a substantial increase.  
When including the first level of transitive dependencies, the risks of being affected by vulnerabilities in small-medium libraries with high technical leverage increase from 4 to 6.7 times, compared to libraries with low technical leverage.
This expanded analysis strongly suggested that transitive dependencies have the potential to substantially increase both opportunities (amount of code borrowed) and risks (vulnerabilities) of technical leverage.

\noindent
In summary, our study delved into three key research questions by scrutinizing the opportunities and risks associated with the direct dependencies of 142 libraries, spanning across a comprehensive dataset of 14,042 releases. This methodological choice was made to uphold the comparability of our findings with those of the replicated study. Subsequently, we extended upon the replicated study by incorporating a broader assessment that encompassed level-1 transitive dependencies, a facet not explored in the initial replication. We then present an in-depth analysis of the implications stemming from the inclusion of level-1 transitive dependencies in all conducted analyses. Table \ref{tab:sim_diff} highlights both the similarities and distinctions between our study and the replicated study.

\begin{table}[]

    \centering
 
    \caption{Unveiling the similarities and differences between our work and the replicated study.}
    
    \begin{adjustbox}{max width=\columnwidth}
\begin{tabular}{lcccccc}
       
        \toprule
        \textbf{} & \textbf{RQ1.Leveraging} &\textbf{RQ2.Benefits}& \textbf{RQ3.Risks} & \textbf{RQ4.Transitive deps} & \textbf{Dataset Size } \\
        &&&&&(packages/releases)\\
      \midrule
        \textbf{Replicated study} & \checkmark & \checkmark & \checkmark & \ding{55}&  {(200/8,494)}   \\
        \textbf{Our study} & \checkmark&\checkmark& \checkmark&\checkmark & {(142/14,042)}\\ 
          
        \hline
\end{tabular}
  \end{adjustbox}
    
    \label{tab:sim_diff}
   
\end{table}

Our study contributes to the research community and practitioners on three fronts:

 \begin{itemize}
     \item To the best of our knowledge, we present the first replication study that evaluates the technical leverage in the NPM ecosystem. We empirically investigated 14,042 NPM library releases to explore the opportunities and risks of direct technical leverage. 
     
     \item We compare our findings with the Maven ecosystem to understand the similarities and differences of opportunities and risks across both ecosystems. 
     Hence. our study helps build a comprehensive body of knowledge on technical leverage in software development. 
     

    \item We complement the replicated study by including a novel analysis considering the first level of transitive dependencies. Transitive dependencies can have significant impact on project development, and  constitute the largest share of dependencies. This expanded analysis help us establish a more realistic view about the benefits and downsides of the technical leverage. 
    
    \item We make our dataset publicly available to foment more research in this area~\cite{dataset_link}. Our dataset is carefully mined and curated, comprising 14,042 stable releases containing metadata about library size, technical leverage, and vulnerability reports.  
 
 \end{itemize}
  
Our findings shed the light on the importance of developers' choices of third-party libraries, since such dependencies are a mixed blessing. This means that reusing other developers' code is an opportunity to increase productivity to an extent, but comes at higher odds of exposing software projects to serious vulnerabilities.

\textbf{Paper Organization.} The paper is organized as follows. 
Section \ref{tech_lev} explains the background and defines the terminology used throughout the study. 
Section \ref{data_selection} describes the case study setup, including the process of collecting and curating our data. 
In Section \ref{study_result}, we dive into our study by answering the four research questions.
The implications of our results to the community  are elaborated in Section~\ref{impl}.
We present the related work in Section \ref{related_work}. 
We state the threats to validity and limitations to our study in Section \ref{threat}. 
Finally, Section \ref{conclusion} concludes our paper.

\section{Background}
\label{tech_lev} 

In this section, we discuss the metrics and terminology used throughout the paper.
Technical leverage (Section~\ref{sub:technical-leverage}) expresses how much of the project's code is borrowed from FOSS, and the evolution metrics (Section~\ref{sub:evolution-metrics}) showcase how projects have evolved with regard to FOSS usage and maintenance.




\subsection{Direct Technical Leverage}
\label{sub:technical-leverage}

The technical leverage is defined as the ratio between the total size of code borrowed from dependencies (both direct and transitive) and the standard libraries, over the size of the project's code. 
As a replication study, we opt to conduct our study on direct technical leverage. 
The \textbf{direct technical leverage ($\lambda$\textsubscript{dir})} is a simplified version of the technical leverage that only considers the code borrowed from direct dependencies from third-party libraries. 
Hence, we exclude from our analyses code borrowed from the standard library and transitive dependencies. 
There are multiple reasons as to why we decided to focus on direct technical leverage:
First, to make our results comparable to the replicated study as the authors also report on the direct technical leverage. 
Second, related work shows that transitive dependencies artificially inflate the problem of vulnerabilities, as transitive dependencies are more abundant than direct dependencies and most vulnerabilities in transitive dependencies do not impact the software projects~\cite{kula2018developers,lauinger2018thou,massacci2021technical,pashchenko2018vulnerable}.
Third, vulnerabilities in direct dependencies can be mitigated by developers, as there is potential for deliberate choice of selecting a new vulnerable direct dependency or keeping vulnerable dependencies outdated~\cite{pashchenko2020qualitative}.


 \vspace{1.1\baselineskip}
 \begin{equation} \label{eqs}
    \lambda\textsubscript{dir} = 
    \frac{
        \tikzmarknode{dir}{\highlight{red}{L\textsubscript{dir}}}}
        {\tikzmarknode{own}{\highlight{blue}{L\textsubscript{own}}}
    } 
\end{equation}

\begin{tikzpicture}[overlay,remember picture,>=stealth,nodes={align=left,inner ysep=1pt},<-]
    \path (dir.north) ++ (0,1em) node[anchor=south east,color=red!67] (dirlabel){size of direct dependencies};
    \draw [color=red!87](dir.north) |- ([xshift=-0.3ex,color=red]dirlabel.south west);
    \path (own.south) ++ (0,-1em) node[anchor=north west,color=blue!67] (ownlabel){size of own code};
    \draw [color=blue!57](own.south) |- ([xshift=-0.3ex,color=blue]ownlabel.south east);
\end{tikzpicture}
\vspace{1.1\baselineskip}

Where L\textsubscript{dir} (dependency code size) represents the sum of the lines of third-party direct dependencies code. L\textsubscript{own} (own code size) is the number of own lines of code in the files of a library, excluding the test files.

\subsection{Evolution Metrics}
\label{sub:evolution-metrics}

The technical leverage metric helps us understand how much of the project code is borrowed from FOSS in a snapshot of a project. 
However, if we want to investigate the evolution of FOSS usage and maintenance over time, we need to introduce the following evolution metrics: change distance and change direction \cite{massacci2021technical}. 

\textbf{Change distance ($\rho$).}
The change distance characterizes the amount of a change in the code size between two consecutive library releases $r0$ and $r1$. In other words, it quantifies the impact of code changes on the library releases time using the following equation: 

\vspace{\baselineskip}
\begin{equation} \label{eq2}
\rho  = \sqrt {
    \tikzmarknode{own}{\highlight{blue}{\Delta L\textsuperscript{2}\textsubscript{own}}} + 
    \tikzmarknode{dir}{\highlight{red}{\Delta L\textsuperscript{2}\textsubscript{dir}}}
    }
\end{equation}
\begin{tikzpicture}[overlay,remember picture,>=stealth,nodes={align=left,inner ysep=1pt},<-]
    \path (dir.north) ++ (0,1em) node[anchor=south east,color=red!67] (dirlabel){\textbf{$\Delta L\textsubscript{dir}=  L\textsubscript{dir} (r_{1}) - L\textsubscript{dir} (r_{0})$}};
    \draw [color=red!87](dir.north) |- ([xshift=-0.3ex,color=red]dirlabel.south west);
    \path (own.south) ++ (0,-1em) node[anchor=north west,color=blue!67] (ownlabel){\textbf{$\Delta L\textsubscript{own}=  L\textsubscript{own} (r_{1}) - L\textsubscript{own} (r_{0})$}};
    \draw [color=blue!57](own.south) |- ([xshift=-0.3ex,color=blue]ownlabel.south east);
\end{tikzpicture}
\vspace{\baselineskip}

\vspace{\baselineskip}



\textbf{Change direction ($\theta$).}
The change direction characterizes the evolution type of a library between two consecutive releases $r_{0}$ and $r_{1}$ by quantifying how developers change their own code and their dependencies using the following equation:


\vspace{\baselineskip}
\begin{equation} \label{eq3}
\theta = \arccos\left(\frac{
    \tikzmarknode{ldir}{\highlight{red}{\Delta L\textsubscript{dir}}}
}{
    \tikzmarknode{rho}{\highlight{blue}{\rho}}
}\right)*
\begin{cases} 
   +1 & \text{if } \Delta L\textsubscript{own} > 0 \\
   -1 & \text{if } \Delta L\textsubscript{own} \leq 0
\end{cases}
\end{equation}

\begin{tikzpicture}[overlay,remember picture,>=stealth,nodes={align=left,inner ysep=1pt},<-]
    \path (ldir.north) ++ (5,1em) node[anchor=south east,color=red!67] (ldirlabel){\textbf{$\Delta L\textsubscript{dir}=  L\textsubscript{dir} (r_{1})- L\textsubscript{dir} (r_{0})$}};
    \draw [color=red!87](ldir.north) |- ([xshift=-0.3ex,color=red] ldirlabel.south east);
    \path (rho.south) ++ (0,-1em) node[anchor=north west,color=blue!67] (rholabel){change distance};
    \draw [color=blue!57](rho.south) |- ([xshift=-0.3ex,color=blue] rholabel.south east);
\end{tikzpicture}
\vspace{\baselineskip}
\vspace{\baselineskip}

\begin{figure}
\includegraphics[width=.7\linewidth, trim=1cm 13.5cm 2.5cm 4cm , clip]{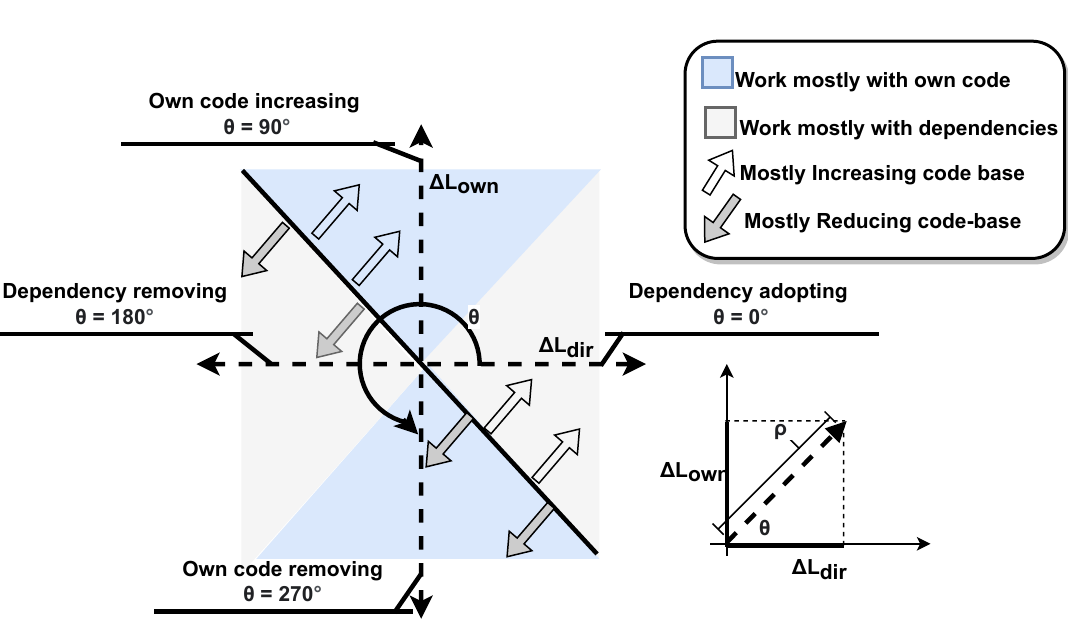}
\caption{Representation of the library change types according to the metrics of change distance ($\rho$) and change direction ($\theta$), proposed by the replicated study~\cite{massacci2021technical}. }
\label{coord}
\end{figure}


Figure~\ref{coord} illustrates the change direction ($\theta$) on a coordinate plane. The x-axis and y-axis of a plane represent the change size of own code and dependency code, respectively. The change direction ($\theta$) captures the polar coordinates of the changes in the plane between the own and dependency code sizes as shown in Figure~\ref{coord}. This coordinate plane elaborates the strategies used by developers to evolve their libraries, such as adding/removing dependencies. The following change directions illustrate the basic concepts of different types of library evolution as depicted in Figure~\ref{coord}:

\begin{itemize}
    \item If $\theta$ $\approx$ 0$\degree$, means, developers increase the size of library dependencies, but do not change their library own size,~$\Delta$L\textsubscript{dir} $>$ 0~ and~ $\Delta$L\textsubscript{own} $\to$ 0.
    
    \item If $\theta$ $\approx$ 90$\degree$, means, developers do not change the library dependency size, but increase their library own code size, $\Delta$L\textsubscript{dir}$\to$ 0~ and~ \(\Delta\)L\textsubscript{own}$>$ 0.
    \item If $\theta$ $\approx$ 180$\degree$, means, developers decrease the library dependency size, but do not change their library own code size, $\Delta$L\textsubscript{dir}$<$ 0~ and~ $\Delta$L\textsubscript{own} $\to$ 0.
    \item If $\theta$ $\approx$ 270$\degree$, means, developers decrease the library's own code size, but do not change their library dependency size, $\Delta$L\textsubscript{dir}$\to$ 0~ and~ $\Delta$L\textsubscript{own}$<$ 0.
\end{itemize}

A combination of these library evolution directions ($\theta$) captures every library change. For example, if \(\theta\) \(\in\) (0\(\degree\), 90\(\degree\)) for a library, this means that developers adopt new dependencies and at the same time change the library own code (i.e., make some development).

\section{Case Study Setup}
\label{data_selection}
The main goal of this paper is to study how technical leverage is associated with opportunities for faster release cycles and risks of security vulnerabilities.
In this section, we describe the data collection used in the rest of our study (Section~\ref{sub:collecting-NPM-libraries}), explain the calculation of NPM libraries and their respective releases sizes (Section~\ref{sub:calculate_size}), and the collection of their associated vulnerability reports (Section~\ref{sub:collecting_vulerabilities}).

\subsection{Collecting NPM libraries}
\label{sub:collecting-NPM-libraries}

We aim to conduct our study on very popular NPM libraries, as these libraries tend to be well-maintained and have the tendency to impact the ecosystem at large \cite{zerouali2019impact,abdalkareem2020impact,zapata2018towards,moller2020detecting,nielsen2021semantic}.
We start our methodology by collecting NPM libraries using the libraries.io database~\cite{andrew_nesbitt_2017_808273,Libraries}.
The libraries.io is an open source repository and dependency database that catalogues libraries of the most popular ecosystems (e.g., NPM), and it has been used by previous work as a source of library metadata~\cite{zerouali2019diversity,decan2019empirical,decan2018impact,husain2019codesearchnet}. 
To identify the most popular packages in NPM, we select the libraries with the highest number of dependent packages as a proxy of library popularity, similarly as done in prior work~\cite{zapata2018towards,zimmermann2019small,zerouali2019diversity}.
To compare our results with the ones in the replicated study \cite{massacci2021technical}, we select a number of libraries that would yield a comparable number of library releases. 
The replicated study investigated 200 Java libraries accounting for 8,464 library releases. However, NPM packages have a higher frequency of releases compared to Maven libraries~\cite{imtiaz2021comparative,Libraries}.
Therefore, we select the most popular 142 libraries, which yields a total of 24,820 library releases.

We opted for a smaller dataset of libraries primarily due to the substantially higher release frequency observed within the NPM ecosystem. This approach (1)  allows us to focus more closely on the trends and patterns utilized by developers within individual libraries, thereby facilitating a deeper understanding of their usage and impact, (2)  enables us to establish comparable conditions with the replicated work. While the replicated study features 40\% more packages than the 142 in our dataset as elaborated in Table \ref{desc_stat}, the number of filtered releases in our dataset is 66\% higher than in the replicated study. Thus, we decided to select fewer packages to keep the analysis more manageable. Moreover, if we had included more packages, we would have significantly increased the number of transitive dependencies, making it impossible to conduct the carried analysis.

In our study, we also want to evaluate how technical leverage is associated with the release cycle of a library. Hence, we need to establish a mature linear evolution of libraries based on their release. To achieve this, we filter out experimental releases, which are unstable releases discouraged from being reused by the community. We use a regular expression to exclude experimental releases if they are tagged as ‘experimental.’
Moreover, we exclude backport releases by organizing releases in chronological order according to their release timestamps. Subsequently, we developed a script to automate the detection of backports, filtering out those versions from consideration based on their timing order. To enhance accuracy, we utilized Semantic Versioning (SemVer) to validate whether a version qualifies as a backport, considering scenarios where changes from a higher minor or patch version are applied to a lower one.
For instance, consider a software project following Semantic Versioning (SemVer) with the following release history:\\
- v1.3.0 (released on 2024-01-01)\\
- v1.3.1 (released on 2024-01-15)\\
- v1.2.3 (released on 2024-02-01)\\
In this example, v1.2.3 was released after v1.3.1. Using SemVer, we determined that v1.2.3 qualifies as a backport since its changes pertain to the v1.2.x branch, even though a more recent minor version, v1.3.x, exists. Our script verifies this by cross-referencing release timestamps and version identifiers, confirming that v1.2.3 introduces fixes or updates backported from higher versions (e.g., v1.3.x).

\begin{table}[ht!]
\caption{Descriptive statistics of Maven (from the study of Massacci and Pachenko~\cite{massacci2021technical}) and the selected NPM libraries of our study. In the NPM libraries, we consider a 10KLOC threshold to differentiate small-to-medium-sized libraries from large-sized libraries.}
\centering
\begin{tabular}{l|lrrrrrrr}

\toprule

\textbf{Maven Libraries} & 
\textbf{Item} & 
\textbf{Mean} & 
\textbf{Min} & 
\textbf{Median}& 
\textbf{Max}\\ 
\midrule

& \# Versions per library &
 55 &
 1 &
 35&
 248 \\
 
\textbf{All} & \# Direct dependencies &
 4&
 0&
 2&
 51 \\
 
 & L\textsubscript{own}(KLOC) &
  37&
2&
15&
350
 \\ 
 
& L\textsubscript{dir}(KLOC) &
 591&
 0&
 302&
 4,489\\ 

 \\

\toprule
\textbf{NPM Libraries} & 
\textbf{Item} & 
\textbf{Mean} & 
\textbf{Min} & 
\textbf{Median}& 
\textbf{Max}\\ 

\midrule
 
& \# Versions per library &
 100 &
 9 &
 67&
 432 \\
 
& \# Direct dependencies &
 7&
 0&
 3&
 83 \\
 
 & L\textsubscript{own}(LOC) &
  10,380
&
  10&
  2,042
&
  204,985
 
 \\ 
 
& L\textsubscript{dir}(LOC) &
 20,608&
 0&
 596&
715,849\\ 

\textbf{All} &\# Level-1 trans. dep. &
 16&
 0&
 4&
 320 \\
 
& L\textsubscript{level-1 trans. dep.}(LOC) &
 20,744&
 0&
 1,245&
 422,606\\ 

& Release interval (days)&
 23&
 0&
 6&
 1,716 \\ 
& \# Files &
 112 &
 1 &
 15&
 34,543 \\
 
& \# Functions &
 1,402 &
 1 &
 214&
228,850 \\ 

\midrule
&\# Direct dependencies &
 7&
 0&
 3&
 83 \\

& L\textsubscript{own}(LOC) &
  2,123
&
  10&
  948
&
  9,991

 \\ 
 
& L\textsubscript{dir}(LOC) &
11,893&
 0&
 2,428&
 393,508\\ 

\textbf{Small-medium}&\# Level-1 trans. dep. &
 18&
 0&
 5&
 320 \\
& L\textsubscript{level-1 trans. dep.}(LOC) &
  23,473&
 0&
 1,835&
 422,606\\ 
& \# Files &
 39 &
 1 &
 10&
 34,543 \\
 
& \# Functions &
446 &
 1 &
94&
228,850 \\

\midrule

&\# Direct dependencies &
 7&
 0&
 1&
 48 \\

& L\textsubscript{own}(LOC) &
  31,963
&
  10,016&
  21,513
&
  204,985

 \\

& L\textsubscript{dir}(LOC) &
20,608&
 0&
  596&
 715,849\\

\textbf{Large }&\# Level-1 trans. dep. &
 11&
 0&
 1&
111 \\

& L\textsubscript{level-1 trans. dep.}(LOC) &
  13,652&
 0&
 184&
 127,301\\ 

& \# Files &
 303 &
 1 &
 97&
6,677 \\
 
& \# Functions &
3,889 &
 1 &
3,150&
35,651 \\

 \bottomrule
 \end{tabular}
\label{desc_stat}
\end{table}



The result of this filtering is a final set that includes 14,042 library releases. Table~\ref{desc_stat} provides comprehensive statistics for the 142 selected NPM  libraries. It includes descriptive statistics for two distinct subgroups of releases (small-medium and large) after excluding backport and experimental releases. The table also presents key metrics related to the number of files and functions in the releases, along with information on level-1 transitive dependencies.
We observe that the selected libraries have numerous releases (median of 67), a low number of direct and level-1 transitive dependencies (median of 3 and 4 respectively), and are typically smaller in size (2,042 LOC). Furthermore, the large libraries exhibit a substantial difference in scale compared to small-medium libraries, with approximately 10 times more files and 33 times more functions.
In our examination of small-medium and large libraries, we discovered that small-medium libraries account for 10,154 releases, while large libraries have 3,888 releases. Additionally, it is evident that small-medium libraries exhibit three times more direct dependencies and five times more level-1 transitive dependencies compared to large libraries.
Our dataset includes the most dependent upon libraries in NPM, such as eslint (280k dependents), mocha (236k dependents), and lodash (150k dependents).

In the replicated study, the authors started from the top 200 FOSS Maven-based libraries used by a large software manufacturer across over 500 Java projects. The resulting set corresponds to 8,464 library versions, which in turn include widely used libraries. In our case, we selected the most popular 142 libraries, which yields a total of 14,042 library releases, after filtering. NPM packages have a higher frequency of releases compared to Maven libraries \cite{imtiaz2021comparative}, for example, the statistics in Table \ref{desc_stat} showed that the median number of package releases in NPM is 67 which is about 1.9 times the number of package releases reported in the dataset of maven. Both datasets showed that the median number of direct dependencies are 3 (npm) and 2 (maven). And the median size of the own code is about 7 times larger in maven.This information could potentially impact the technical leverage. For instance, the median size of code being larger in Maven libraries suggests potentially more comprehensive or complex libraries. This could impact technical leverage by offering more functionality out-of-the-box, potentially reducing the need for custom code development. 
Moreover, the Maven study opted for a threshold of 100KLOC to differentiate between small-medium size libraries and large size libraries. We acknowledge that varying thresholds could yield different outcomes. Thus, we investigated alternative thresholds (i.e., 5, 10, 15, and 20 KLOC) and observed that our final findings remain consistent across these different thresholds.
\subsection{Collecting and resolving direct and level-1 transitive dependencies.}
In this section, we describe how we automate the dependency collection and resolution mechanism. The algorithm starts by collecting the direct dependencies specified in a specific package's configuration file (e.g., package.json). Then, for each direct dependency, the algorithm queries the npm registry to obtain information about the available historical versions and their dependencies. After that, the algorithm iterates through the historical versions of the direct dependency starting from the first created version and stopping at the version (target)  created or modified closest to the specified package time. During this iteration, the algorithm checks if the target version satisfies the semantic version of the non-resolved version, using semver.js library. If a satisfying version is found, it updates the resolved version with the target version. For each resolved direct dependency, the algorithm repeats the process recursively for their level-1 transitive dependencies. Our process ensures that our dependencies are resolved as if they were installed at the time of the release of the library project, minimizing the chances of inaccuracies caused by simple reachability analysis algorithms \cite{liu2022demystifying}.
For example, if we have a package webpack@0.4.1 created at 14-05-2012 and one of its direct dependencies is  'esprima', the algorithm iterates through the historical versions of the dependency (esprima) , starting from the first created version and stopping at the version (target)  created or modified closest to the specified package time (webpack@0.4.1).
During this iteration, the algorithm checks if the target version satisfies the semantic version of the non-resolved version, using semver.js library. The  non-resolved version collected from the package.json of the package (webpack@0.4.1). If a satisfying version is found, it updates the resolved version with the target version.

\subsection{Calculating the size of libraries and their dependencies}
\label{sub:calculate_size}

To compute the direct technical leverage, we calculate the library's own size and the sum of all its direct dependencies' size metrics. 
We measure the size in lines of code using the CLOC tool\footnote{https://github.com/AlDanial/cLOC}.
CLOC is a utility program that helps us count the size of a library or dependency without accounting for comments and blank lines. 
As direct technical leverage focuses only on the size of the actual source code of a library, we exclude test files from our analysis, for both own and dependency code, by configuring CLOC using the {\small\texttt{--exclude-dir}} parameter. We depend on the common and recommended structure for a JavaScript package which often follows a modular approach to remove all test directories. To eliminate testing code, we systematically removed directories named test, tests, testing, \textunderscore\textunderscore test \textunderscore\textunderscore, or spec  by configuring CLOC using the --exclude-dir parameter, as well as files named test.js, tests.js, testing.js, test-*.js, *-test.js, or *.spec.js.
To calculate the size of library dependencies, we compute the size of each dependency using the CLOC tool and then sum the resulting number of lines of code. It is important to note that we only count JavaScript lines and exclude all other programming language's lines for the own and dependency code.

We employ the same process to calculate the size of all releases and their dependencies described in Section~\ref{sub:collecting-NPM-libraries}. 
Table \ref{desc_stat} shows the statistics of lines of code of the library and their respective dependencies in our dataset. 
We notice that the size of dependencies (L\textsubscript{dir}) is larger than the size of library's own code (L\textsubscript{own}). 
This shows that, on median, the most depended upon NPM libraries leverage more than their own code in FOSS. 

\subsection{Collecting vulnerability data}
\label{sub:collecting_vulerabilities}

To study the risks as a byproduct of direct technical leverage, we measure the number of vulnerabilities reported in the library's own code versus the vulnerabilities reported on their respective direct dependencies.
For this purpose, we mine the Snyk database~\cite{snyk_database}. 
Snyk is an open source security platform for finding out vulnerabilities in different ecosystems (e.g., NPM, Maven, PyPI, and Go) and has been used in prior work to identify whether the code of a library is affected by vulnerabilities~\cite{massacci2021technical,chinthanet2021lags,zerouali2019impact}.
We collected 3,176 vulnerabilities on the NPM libraries and their dependencies on the 1\textsuperscript{st} September 2021. 
An example of the information collected for each vulnerability is shown in Table~\ref{stat_meta_vul}.


\begin{table}
\caption{Example of the collected metadata for vulnerabilities.}
\centering

\begin{tabular}{ll}
\toprule

\textbf{Metadata} & \textbf{Value} \\

\midrule

\textbf{Vulnerability Name} &  Command Injection \\
\textbf{Affected Packages} & Codecov \\
\textbf{Affected Versions} & $<$ 3.7.1 \\
\textbf{Severity} & Medium \\
\textbf{Published Date} & 21 Jul, 2020 \\

\bottomrule

 

 \end{tabular}
\label{stat_meta_vul}
\end{table}


\begin{figure}
    \centering
    \includegraphics[width=.8\linewidth]{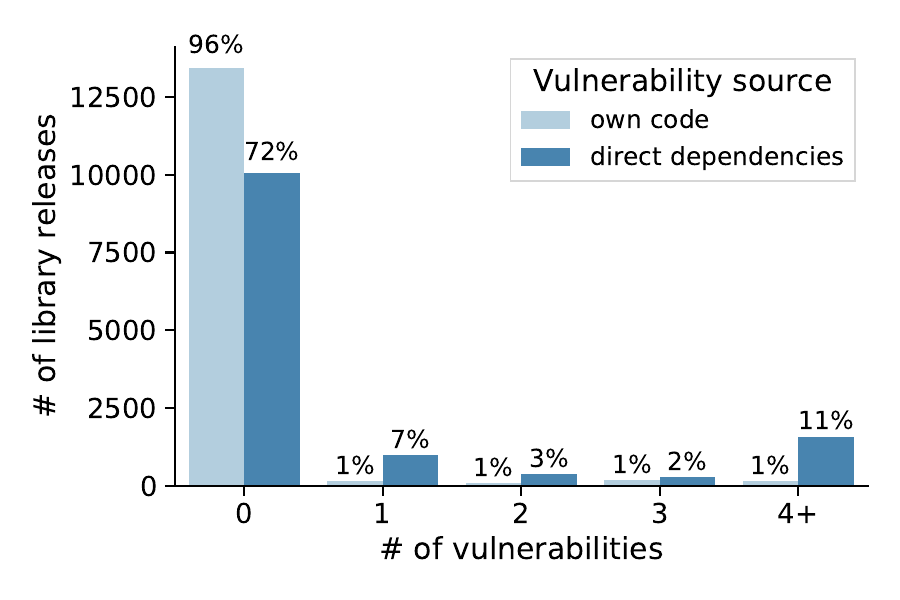}
    \caption{Distribution of releases with reported vulnerabilities from their own code and dependencies in our dataset.}
    \label{fig:vulnerabilities}
\end{figure}



Figure~\ref{fig:vulnerabilities} presents the percentage of releases in our dataset associated with vulnerabilities for both own code and dependencies code. 
From the figure, we observe that 28\% of our dataset has at least one vulnerability stemming from direct dependencies and only 4\% coming from native code developed by project maintainers. Moreover, the figure shows that the percentage of releases suffered from at least one vulnerability is always higher in direct dependencies compared to ones coming from own code. 


\section{Case Study Results}
\label{study_result}

In this section, we present the findings of our study with respect to our research questions. For each research question, we present the motivation for the question, describe the approach used to answer it, and discuss the results of our analysis.

\subsection{RQ1: Is there a difference in direct technical leverage, distance and direction of changes between small-medium and large libraries?}
\label{rq1}

\noindent
\textbf{Motivation:}
The NPM ecosystem is the fastest growing ecosystem to date, with more than 2.15 million libraries available for the community.
Thus, we expect the NPM ecosystem to have a higher degree of direct technical leverage than reported in the Maven study of Massacci and Pashchenko~\cite{massacci2021technical}. 
Therefore, in this RQ, we examine the direct technical leverage, distance, and direction of changes for small-medium and large libraries~\cite{massacci2021technical} in the NPM ecosystem. 
This enables us to understand the different strategies that developers use to evolve and maintain their libraries based on the library size. 
\\

\noindent
\textbf{Approach:}
We first divide our dataset into small-medium and large library releases. 
To achieve this, we choose the third quantile (i.e., Q75) of our sample distribution as a threshold, which is equivalent to 10KLOC, to split the dataset into small-medium ($<$10KLOC) and large library ($\geq$10KLOC) releases.
Then, we compute the direct technical leverage ($\lambda$$\textsubscript{dir}$) on both small-medium and large library releases and compare their distributions. 

To evaluate how developers change their code for small-medium and large libraries, we compute the change distance (\(\rho\)) and change direction (\(\theta\)) metrics for both sets. 
As these metrics are computed across releases, for each release $r_{1}$ in our dataset, we use the closest previous release as $r_{0}$ to calculate them as described in Section~\ref{sub:evolution-metrics}.
Then we use the kernel density estimation (KDE) to plot the distribution of change direction in both sets. 
KDE is closely related to histograms but can be favoured with properties such as smoothness or continuity by using an appropriate kernel.
\\

\begin{table}
\caption{Descriptive statistics of the metrics.}
\centering
\begin{tabular}{p{0.55in}rrrrrr}
\toprule

\textbf{Item} & \textbf{Mean}  & \textbf{Min} & \textbf{Q25\%} &  \textbf{Median}& \textbf{Q75\%} & \textbf{Max}\\ 

\midrule
 \(\lambda\textsubscript{dir\textunderscore{small-med-lib}}\)&
 36&
 0&
 0.163&
 2.53&17&
 7,709\\ 

 \(\lambda\textsubscript{dir\textunderscore{large-lib}}\) &
 1.2&
 0&
 0.005&
 0.034&
 0.64&
58\\ 

 \(\rho (LoC)\) &
 2,096&
 0&2&
 25&
 249&
 647,106\\ 
 \(\theta (degree)\) &
 1.5&
 -180&
 -90&
 0.5&
 90&
 180 \\ 
 \bottomrule
 \end{tabular}
\label{stat_met}
\end{table}


\begin{figure*}
\includegraphics[width=\linewidth]{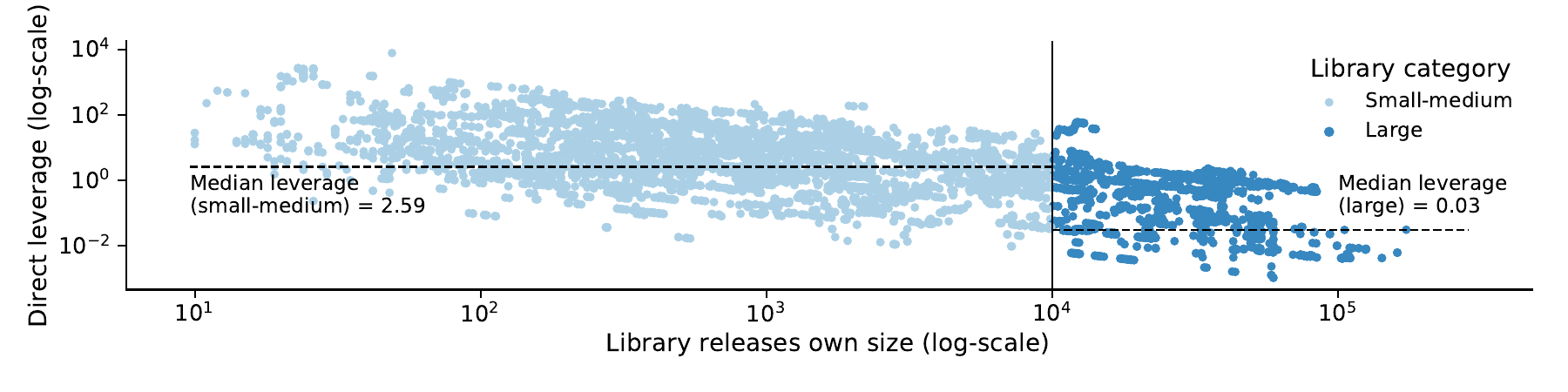}
\caption{The direct technical leverage of 14,042 library releases per library size in LOC.}
\label{direct_lev}
\end{figure*}



\noindent
\textbf{Results:} 
Table~\ref{stat_met} shows the descriptive statistics of the code metrics used to measure the evolution of one's own code and third-party code into a software library. 
Small-medium library releases leverage, on median, 253\% (2.5 times) of their own code in FOSS, while large library releases leverage only 3\% of their code as in FOSS. 
In other words, developers widely adopt dependencies, especially for small-medium libraries, as using dependencies may help small libraries grow quickly in their code and reduce their development effort and time~\cite{grinter1996understanding,mohagheghi2007quality}.


To better understand the relation between the own code size of the libraries (small-medium and large) and the direct technical leverage, we present Figure \ref{direct_lev}. 
In this figure, we present all the 14,042 library releases that fall in the two-dimensional space of technical leverage (y-axis) and their library size (x-axis).
From the figure, we observe that releases that ship more own's code (right part of the graph) also leverage proportionally less code from dependencies. 
Naturally, as direct technical leverage considers the size of a project, the larger the size of its own code, the smaller the likelihood of high technical leverage. 
We confirm this observation by calculating the Spearman Rank Correlation Coefficient \cite{zar2005spearman} and finding a negative linear correlation between direct technical leverage and the size of own code ($\rho$ = -0.466; p-value =0). This means that releases that include a higher proportion of proprietary code tend to rely proportionally less on external dependencies. Moreover, the p-value suggests that the observed negative correlation is statistically significant. Furthermore, we calculated Cohen's d \cite{cohen} to quantify the effect size, resulting in a Cohen's d value of 0.76. This indicates a large association between direct technical leverage and the size of own code.

Still, it is interesting to observe that the leverage of dependencies is not proportional to the size of own's code, as it may indicate a limit to the usefulness of technical leverage as libraries grow.

\begin{figure}

    \begin{subfigure}{.49\linewidth}
        \includegraphics[width=\linewidth]{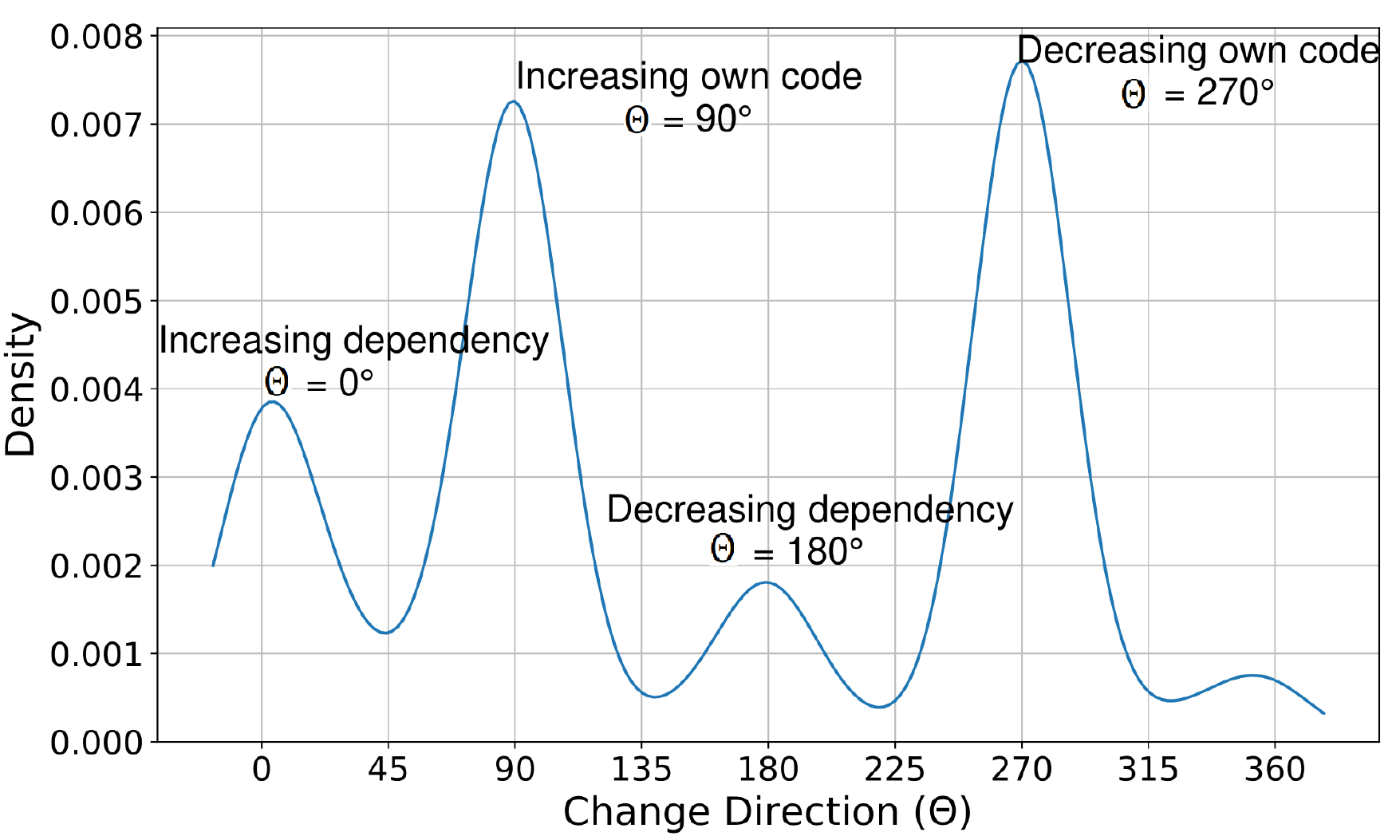}
        \caption{Small-medium library releases ($<$10KLoC).}
        \label{kdesmall}
    \end{subfigure}
    \begin{subfigure}{.49\linewidth}
        \includegraphics[width=\linewidth]{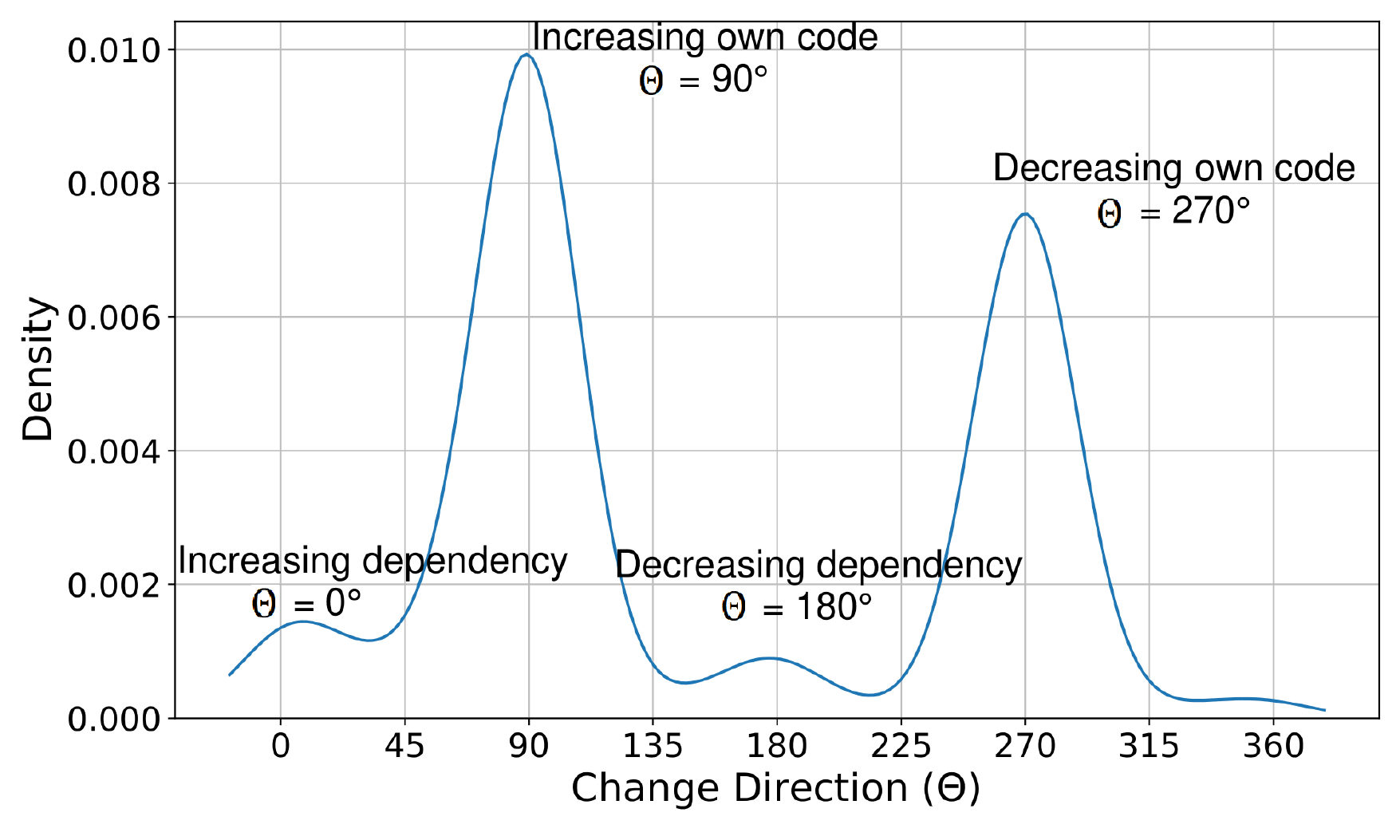}
        \caption{Large library releases ($\geq$10KLoC).}
        \label{kdelarge}
    \end{subfigure}
    
    \caption{Kernel density estimation (KDE) plot with the distribution of the change direction of library releases. }

\end{figure}

\begin{figure}

\end{figure}
Regarding the differences in the type of changes developers perform in their code between small-medium and large libraries, we report on the distribution of change direction across all 14,042 releases.
Figures \ref{kdesmall} and \ref{kdelarge} present the KDE plots for change direction($\theta$) of small-medium and large libraries, respectively. 
A peak in the KDE distribution plots is proportional to the frequency of a type of change across all releases in the dataset. 
From the figures, we observe that developers experience different strategies for updating their small-medium and large libraries. 
In small-medium libraries, developers frequently work in both their own code and adopting new dependencies. 
Figure~\ref{kdesmall} suggests that small-medium library developers focus first increasing and decreasing their own code ($\theta = 90\degree$ and $\theta = 270\degree$), but also frequently include more dependencies in their code-base ($\theta = 0\degree$).
This finding indicates that small-medium library developers constantly resort to adding new dependencies to incorporate more features and reduce development time~\cite{grinter1996understanding,mohagheghi2007quality}. 
On the other hand, in large libraries, developers tend to mostly work on their own code. This is clearly shown in Figure~\ref{kdelarge} where the KDE plot for \(\theta\) = 0$\degree$ has a much smaller curve compared to \(\theta\) = 90$\degree$. 
In other words, developers of large libraries focus on developing/optimizing their code more than using dependencies, compared to small-medium libraries. Moreover, they focus on fixing bugs and vulnerabilities in order to ship healthy libraries to the user.
\\
\noindent
\textbf{Comparison to the Maven ecosystem.} 
\noindent
The replicated study observed similar findings in the Maven ecosystem. Small-medium library releases leverage much more code from dependencies (14.6) than large library releases (0.4) in Java. As a consequence, developers of small-medium libraries frequently work on both their own code and their dependencies, while developers of large libraries work mostly in their own code-base. Surprisingly, however, the direct leverage of NPM libraries is considerably much smaller than in Maven libraries.  We found that small-medium libraries have a median technical leverage of 2.5 versus 14.6 reported by the replicated study in Java small libraries.  Contrary to our assumption, NPM libraries have lower technical leverage than Maven libraries.

The differences in the technical leverage between Maven and npm may be due to distinct factors. One major factor is the differences in average library sizes. (1) The statistical data from the Maven and npm datasets, as presented in Table \ref{desc_stat}, reveals that the median size of both own code and dependency code is notably larger in Maven. (2) Java programs tend to use a wider vocabulary \cite{abdulkareem2021evaluating} and the structure of the Java language is more verbose \cite{flauzino2018you}. Java code tends to have more words due to its static typing and compile-time checks; thus developers may need to write more code to accomplish certain tasks compared to dynamically typed languages (i.e., JavaScript) where the language itself handles type inference. And this is reflected in the size of the library. (3) Additionally, the differences in the module system could contribute to the variations in library size.  Java has a module system introduced in Java 9 (Java Platform Module System), which allows for more modular projects and thus larger libraries. On the other hand, Node.js has a built-in CommonJS module system and supports ECMAScript modules (ESM) since Node.js 12.  ESM allows for more granular control over dependencies and can help reduce project size by only importing what is necessary \cite{turcotte2022stubbifier}.
Overall, other characteristics may also impact, as previous studies \cite {decan2019empirical,kikas2017structure} have shown that ecosystems differ in their evolution. In the same context, developer behaviors, including coding practices, project size preferences, and risk perceptions, may vary between NPM and Maven communities, impacting their reliance on external dependencies and the scale of code leverage.



\begin{tcolorbox}
\textbf{Key Takeaway:} 
Small-medium libraries leverage 2.5 times more code (in median) from dependencies where developers frequently working in adopting more dependencies. In comparison, within the Maven ecosystem, the median direct leverage of dependencies is around 14.65. Large libraries rely on only 3\% of their code from dependencies, in contrast to Maven's 48\%. Despite the variance, developers primarily engage with their own code bases in both contexts. In conclusion, both studies indicate that small to medium-sized library releases leverage significantly more on code from dependencies compared to larger library releases.  Surprisingly, NPM libraries have considerably lower direct technical leverage than Maven libraries.

\end{tcolorbox}

\subsection{RQ2: How does direct technical leverage impact the time interval between library releases?} \label{rq2}
\noindent
\textbf{Motivation:} 
In RQ1, we found that small-medium libraries leverage more than twice their code from dependencies. Prior work shows that software developers reuse code to reduce time-to-market and increase their productivity~ \cite{lim1994effects,decan2018evolution,zerouali2018empirical}. In this RQ, we want to better understand the relation between technical leverage and the speed of releasing libraries. In particular, we assess the associated impact of technical leverage on the release cycle in the NPM ecosystem. Understanding the opportunities associated with technical leverage helps library developers balance between the opportunities and risks associated with technical leverage based on their library needs.
\\




\noindent
\textbf{Approach:}
Following the replicated study, we build a multivariate linear regression model to capture the impact of direct technical leverage, change distance, and change direction on the time interval between library releases~\cite{anderson1962introduction}. 
Take the release interval $\Delta_{x} = r_{x} - r_{x-1}$, calculated using two consecutive releases $r_{x}$ and $r_{x-1}$, the model captures the association between the release interval $\Delta_{x}$ and multiple metrics (e.g., technical leverage) as follows: 


\vspace{.5cm}
\begin{align*}\label{eq_regression}
    \small
         \tikzmarknode{rc}{\highlight{blue}{log(\Delta_{x} + 1)}} 
         = 1 +  
         \tikzmarknode{prc}{\highlight{blue}{log(\Delta_{x-1} + 1)}} 
         + 
         \tikzmarknode{dir}{\highlight{purple}{log(\lambda\textsubscript{dir})}} + \\
         \tikzmarknode{rho}{\highlight{OliveGreen}{log(\rho)}} 
         \color{Bittersweet} 
         \underbrace{
         \color{black}
         + \tikzmarknode{theta}{\highlight{Bittersweet}{cos(\theta - 45\textsuperscript{o})}} 
         + \tikzmarknode{theta}{\highlight{Bittersweet}{sin(\theta)}}
         }
\end{align*} 
\begin{tikzpicture}[overlay,remember picture,>=stealth,nodes={align=left,inner ysep=1pt},<-]
    \path (rc.north) ++ (0,1em) node[anchor=south,color=blue!67] (rclabel){Release cycle};
    \draw [color=blue!57](rc.north)
    |- ([xshift=-0.3ex,color=blue]rclabel.south);
    
    \path (prc.north) ++ (0,1em) node[anchor=south ,color=blue!67] (prclabel){Previous release cycle};
    \draw [color=blue!57](prc.north) |- ([xshift=-0.3ex,color=blue]prclabel.south);
    
    \path (dir.north) ++ (0,1em) node[anchor=south west ,color=red!67] (dirlabel){Technical\\leverage};
    \draw [color=red!87](dir.north) |- ([xshift=-0.3ex,color=red]dirlabel.south east);
    
    \path (rho.west) ++ (-.5em,-2em) node[anchor=east ,color=OliveGreen!85] (rholabel){Amount of change\\(change distance)};
    \draw [color=OliveGreen!67](rho.west) -| ([xshift=-.3ex,color=green]rholabel.north);
    
    \path (theta.south) ++ (-4em,-1em) node[anchor=north ,color=Bittersweet!85] (thetalabel){Direction of changes: total + own};
\end{tikzpicture}
\vspace{\baselineskip}

We describe each independent variable in our regression function as follows:

\begin{enumerate}

    \item \textit{Previous release cycle.} We include the previous release cycle to capture the effect of release practices used by developers (i.e., daily, weekly release) in the dependent variable. As releases can be published on the same day (i.e., release cycle of zero days), it is necessary to increment $+1$ to keep a valid computation under a log function. 

    \item \textit{Technical leverage.} This is the main variable of our study, as we want to identify the associated impact on technical leverage to the release interval cycle. 
    
    \item \textit{Amount of changes.} We use the log of the change distance $log(\rho)$ to capture the amount of changes (both in own code and dependencies) between releases. Intuitively, releases with many changes tend to take longer to be released. 
    
    \item \textit{Direction of changes (total code).} To consider the direction of changes in the total code size, we use the term $cos(\theta - 45)$. Referring to Figure \ref{coord}, one might notice that there is a diagonal line from  \(\theta\) = -45$\degree$ to  \(\theta\) = 135$\degree$. This line separates the behaviour of increasing total code base from the behaviour of decreasing total code base. By shifting the angle in 45$\degree$, we capture the direction of the change in total code size by using the cosine function. 
    
    \item \textit{Direction of changes (own code).} It is also necessary to differentiate between total code changes and changes in one's own code base, as the latter tends to impact the release cycle more substantially. 
    To capture this phenomenon, we use the $\sin(\theta)$ function to capture the ratio of changes performed in one own's code ($\Delta_{own}$).
\end{enumerate} 
    
\begin{figure}[]
\centering
\includegraphics[width=\linewidth]{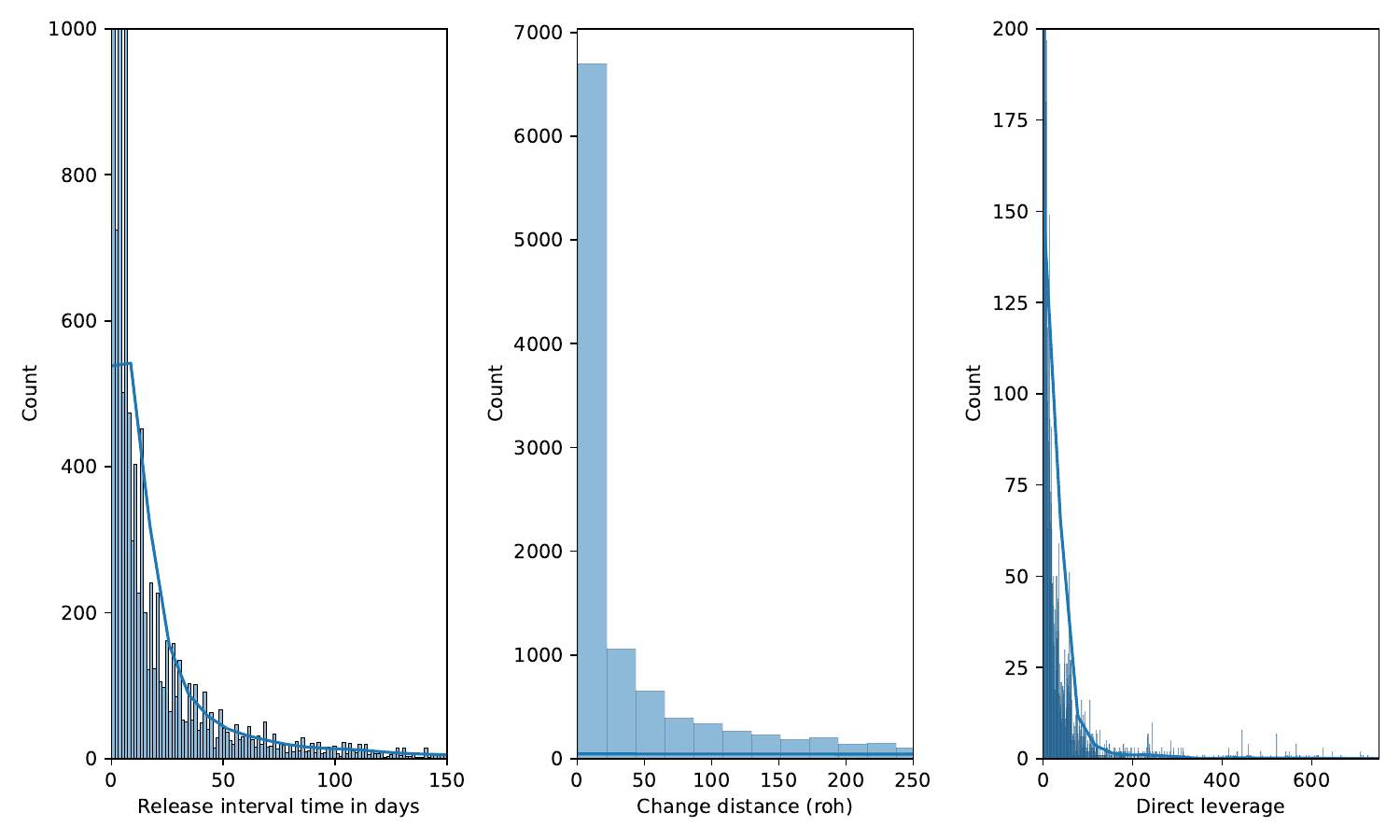}
\caption{Distribution of regression model's variables.}
\label{log-tail}
\end{figure}

We employ a logarithmic scale for the release cycle and technical leverage metrics to effectively represent their pronounced long-tail distributions, as illustrated in Figure~\ref{log-tail}.
Also, we apply the regression model on all releases in our datasets described in Section~\ref{sub:collecting-NPM-libraries} to model the impact of direct technical leverage on the time interval between releases.
Furthermore, we constructed a multivariate linear regression model, encompassing all releases (i.e., global analysis) and we reported the result in Table 5.

\begin{table*}[h]
    \centering
    \caption{Linear regression model fit to check the correlation between \(\lambda\textsubscript{dir}\) , \(\theta\), \(\rho\) and the release time interval.}
    \begin{adjustbox}{max width=\columnwidth}
\begin{tabular}{ll|ll|ll|ll}
        \toprule
        \textbf{Description} &
        \textbf{Coefficients} & 
        \multicolumn{2}{c}{\textbf{Small-medium libraries}} &
        \multicolumn{2}{|c}{\textbf{Large libraries}} &
        \multicolumn{2}{|c}{\textbf{All libraries}} \\
        
         &
         & 
         \textbf{Estimate} &
         \textbf{P-value} &
         \textbf{Estimate} &
         \textbf{P-value} &
         \textbf{Estimate} &
         \textbf{P-value} \\
         
         \midrule
         
         Intercept& 
         1& 
         0.4377  &
         0.000 &
         0.4166 &
         0.000 &
          0.4245&
          0.000
         \\ 
         
         Direct leverage &
         log(\(\lambda\)\textsubscript{dir}) &
         -0.0386   &
         0.000 &
         \textbf{0.0140} &
         \textbf{0.169 }&
        \textbf{0.0050} &
         \textbf{0.329}  
         \\
         
         Change distance &
         log(\(\rho\)) &
         0.1101 &
         0.000 &
         0.0802 &
         0.000 &
         0.0896       & 
         0.000  \\
         
         Change in total code  &
         cos(\(\theta\) - 45\textsuperscript{o}) &
         0.2686 &
         0.000 &
         0.0993  & 
         0.003 &
         0.2456    &
         0.000 \\

         Change in own code&
         sin(\(\theta\)) &
         -0.0578  &
         0.000 &
         \textbf {0.0028}   &
         \textbf {0.914}&
         -0.0594        &   0.000   \\
         
         Previous release interval &
         $log(\Delta_{x-1} + 1)$ &
         0.2640 &
         0.000&
         0.2114     &
         0.000  &
         0.2526      &
         0.000 
\end{tabular}

\end{adjustbox}

\footnotesize
\begin{flushleft}
For small-medium libraries: Root mean square error(rmse) =0.64, R-squared( R\textsuperscript{2}) = 0.181, Adj. R-squared ($\overline{R}$\textsuperscript{2}) = 0.180\\
For large libraries: Root mean square error(rmse) =0.56, R-squared( R\textsuperscript{2}) = 0.124, Adj. R-squared  ($\overline{R}$\textsuperscript{2}) = 0.123 

For all libraries: Root mean square error(rmse) =0.623, R-squared( R\textsuperscript{2}) = 0.158, Adj. R-squared  ($\overline{R}$\textsuperscript{2}) = 0.158
\end{flushleft}

\label{table_reg}
\end{table*}



\noindent
\textbf{Results:}
Table \ref{table_reg} shows the coefficients of our model as well as their respective significance to the model's predictive power. 
According to our model, all metrics show statistical significance when predicting the release interval for small-medium libraries (p-value $<$ 0.05).
The model suggests a small, but negative association between direct technical leverage in release interval ($\lambda\textsubscript{dir} = -0.038$). 
\textbf{In other words, the model suggests that libraries that leverage more code from third-party libraries tend to have shorter release cycles.}
To understand the practical implications of the technical leverage coefficient of the model, let us walk through a hypothetical scenario. 
According to the model, a 4 times increase of technical leverage in libraries would shorten the predicted release cycle by one day (keeping all other metrics unchanged).  
We calculate this by using the data shown in Table~\ref{desc_stat}. 
Using the model, a 4 times increase in the technical leverage of all libraries affects the predicted mean release interval cycle in one day, from 23 days to 22 days, on average.

In large libraries, however, we observe no influence of direct technical leverage on the release interval (p-value $>=$ $0.05$ in Table~\ref{table_reg}). 
This result is inline with the observed characteristics for the large libraries presented in Figure \ref{direct_lev}. \textbf{In other words, large libraries leverage just a small part of their code in third-party dependencies (median of 3\%), the reliance on dependencies code seems to have no significant impact on the libraries release interval.} 
Instead, large libraries release interval seems to be mostly influenced by the project release periodicity (previous release interval), and the change distance of the project's total code ($\rho$ and $cos(\theta - 45\degree)$).

Upon considering the inclusion of all releases (global analysis, both small and large libraries), our analysis results remain consistent with those observed in small-medium libraries. With the exception of direct leverage, we find no significant influence of direct technical leverage on the release interval (p-value = 0.32, as indicated in Table~\ref{table_reg}).
\textbf{This suggests that the release interval of libraries is primarily influenced by the project's release periodicity and the change distance of the project's overall codebase.}

\textbf{Comparison to the Maven ecosystem.} 
In the original study, the authors report a positive coefficient when discussing the impact of technical leverage on release interval. 
That is, shipping more code from FOSS (higher technical leverage) does incur an overhead in the release interval. 
In practical terms, while in small NPM libraries we simulate that a 4 times higher technical leverage would in average shorten the release interval by one day. On the other hand, in Maven, the release interval would be increased by 2-6 days, on average. 
In both ecosystems, direct technical leverage provides an opportunity for better software productivity, with popular libraries being able to ship multiple times its own code with little to no overhead in the release cycle.   

\begin{tcolorbox}
\textbf{Key Takeaway: } In NPM, smaller libraries seem to benefit from faster releases when using more FOSS code, while larger libraries show no significant influence on release cycles from direct technical leverage. In Maven, direct leverage also appears advantageous, as projects leverage more code without significantly affecting the release time. 
\end{tcolorbox}

\subsection{RQ3: Does technical leverage impact the risk of including more vulnerabilities?} \label{security-risk} \label{rq3}

\noindent
\textbf{Motivation:} 
We observe in RQ2 that technical leverage appears to have a measurable positive impact on speeding-up releases.
However, leveraging other people's code comes with its own risks.
As vulnerabilities in NPM libraries are widespread~\cite{decan2018impact,Alfadel21Dependabot}, we expect projects with higher technical leverage to be at a higher risk of being affected by vulnerabilities. 
Hence, in this RQ, we evaluate this hypothesis by assessing the risks of projects with low and high technical leverage to vulnerability exposure.
\\

\noindent
\textbf{Approach:}
To assess the risks associated with direct technical leverage, we use the Odds Ratio (OR) to understand the impact of direct technical leverage on the security risk. 
The OR is commonly used in medicine to assess the effect of a parameter on a rare disease~\cite{szumilas2010explaining}.
We use OR to measure the association between an exposure (high direct leverage) and outcome (security risk), to evaluate if direct leverage is a risk factor for the occurrence of vulnerabilities.

We compare the difference in the associated risk on 1) libraries with low direct technical leverage and 2) libraries with high direct technical leverage. We use the following two definitions in the computation of OR, equation~\ref{or}, as LowLeverageLibs and HighLeverageLibs:

\begin{itemize}
    \item \textbf{Low direct technical leverage (LowLeverageLibs).} Libraries with direct technical leverage that are lower than the medians of small-medium   (\(\lambda\)\textsubscript{dir(small-medium)}$\leq$ 2.5)   and large (\(\lambda\)\textsubscript{dir(large)} $\leq$ 0.034) libraries.

    \item \textbf{High direct technical leverage (HighLeverageLibs).} Libraries with direct technical leverage that are higher than the medians of small-medium (\(\lambda\)\textsubscript{dir(small-medium)}$>$ 2.5) and large (\(\lambda\)\textsubscript{dir(large)}$>$ 0.034) libraries.
    
\end{itemize}

For the risk assessment, we consider a library release to be vulnerable if one or more vulnerabilities are reported in their code or their direct dependencies.
Hence, $VulnLibs$ include all releases that have at least one vulnerability in their code or dependencies, while $NotVulnLibs$ are library releases that are totally free of reported vulnerabilities. 

To compute the OR, we use the following equation:


\begin{equation}\label{or}
\begin{split}
    Odds~ Ratio~(OR) =\frac{\frac{|HighLeverageLibs  ~\cap ~ VulnLibs|}{|HighLeverageLibs ~\cap ~ NotVulnLibs|}}{\frac{|LowLeverageLibs ~ \cap ~ VulnLibs|}{|LowLeverageLibs  ~\cap ~ NotVulnLibs|}}  \\
\end{split}
\end{equation}

The interpretation of the OR value is as follows: 
\begin{itemize}
    \item OR = 1 indicates no security risks associated with the direct technical leverage.
    \item OR $>$ 1 indicates higher security risks (higher odds) associated with the direct technical leverage.
    \item and OR $<$ 1 indicates lower security risks (lower odds) associated with the direct technical leverage.
\end{itemize}

To ensure that the difference between high and low direct technical leverage is statistically significant, we apply the Fisher Exact test\cite{fisher1992statistical}. 
We selected the Fisher Exact test because it is commonly used to determine if there are non-random associations between two categorical variables \cite{cogo2021empirical,zampetti2019study}, and to make our analysis comparable to the replicated study. 
\\

\begin{table}[tb]
\caption{Contingency table for vulnerable libraries vs direct technical leverage with the corresponding Odds Ratio (OR).}
\centering
\begin{tabular}{cccccc}
\toprule
\multicolumn{3}{c}{\textbf{Small-medium libraries}}&
\multicolumn{3}{c}{\textbf{Large libraries}}\\
    & \textbf{\#Vuln}&  \textbf{\#Not vuln} &
    &\textbf{\#Vuln}&  \textbf{\#Not vuln}\\
  \midrule
\textbf{\(\lambda\)\textsubscript{dir} $>$ 2.5} &  3,212&1,918 &\textbf{\(\lambda\)\textsubscript{dir} $>$ 0.034} & 1,251&692\\
\textbf{\(\lambda\)\textsubscript{dir} $\leq$ 2.5}  &1,456&3,588 &\textbf{\(\lambda\)\textsubscript{dir}  $\leq$ 0.034} &368& 1,557\\
\midrule
\textbf{OR} & 4\threeStars & & \textbf{OR} & 7.6\threeStars & \\

\bottomrule
 \end{tabular}
\footnotesize
\begin{flushleft}
\quad \quad  \quad \quad \quad  \quad  \oneStar$p < 0.05$, \twoStars $p < 0.01$, \threeStars$p < 0.001$  \\
\end{flushleft}
\label{table_contengency}
\end{table}


\noindent
\textbf{Results:}
Table \ref{table_contengency} presents the contingency table for vulnerable libraries and direct technical leverage used to calculate the OR. 
\textbf{The results confirm the intuition that high direct leverage helps expose library releases to more vulnerabilities.}
More specifically, we find that the OR for small-medium libraries to be 4.0, indicating that small-medium libraries with high direct leverage ($\lambda_{dir} > 2.5$) have 4 times the chance of being exposed to vulnerabilities than libraries with low direct leverage ($\lambda_{dir} <= 2.5$). 
While large libraries leverage proportionally less code as discussed in RQ1, large libraries that leverage more than 3\% of their own code from dependencies have 7.6 times higher chance of being exposed to vulnerabilities than large libraries with low direct leverage.  
In both cases, we find the differences between high and low direct technical leverage libraries to be statistically significant (p $<$ 0.001). 
 



\begin{figure}[tb]
\centering
\includegraphics[width=.9\linewidth]{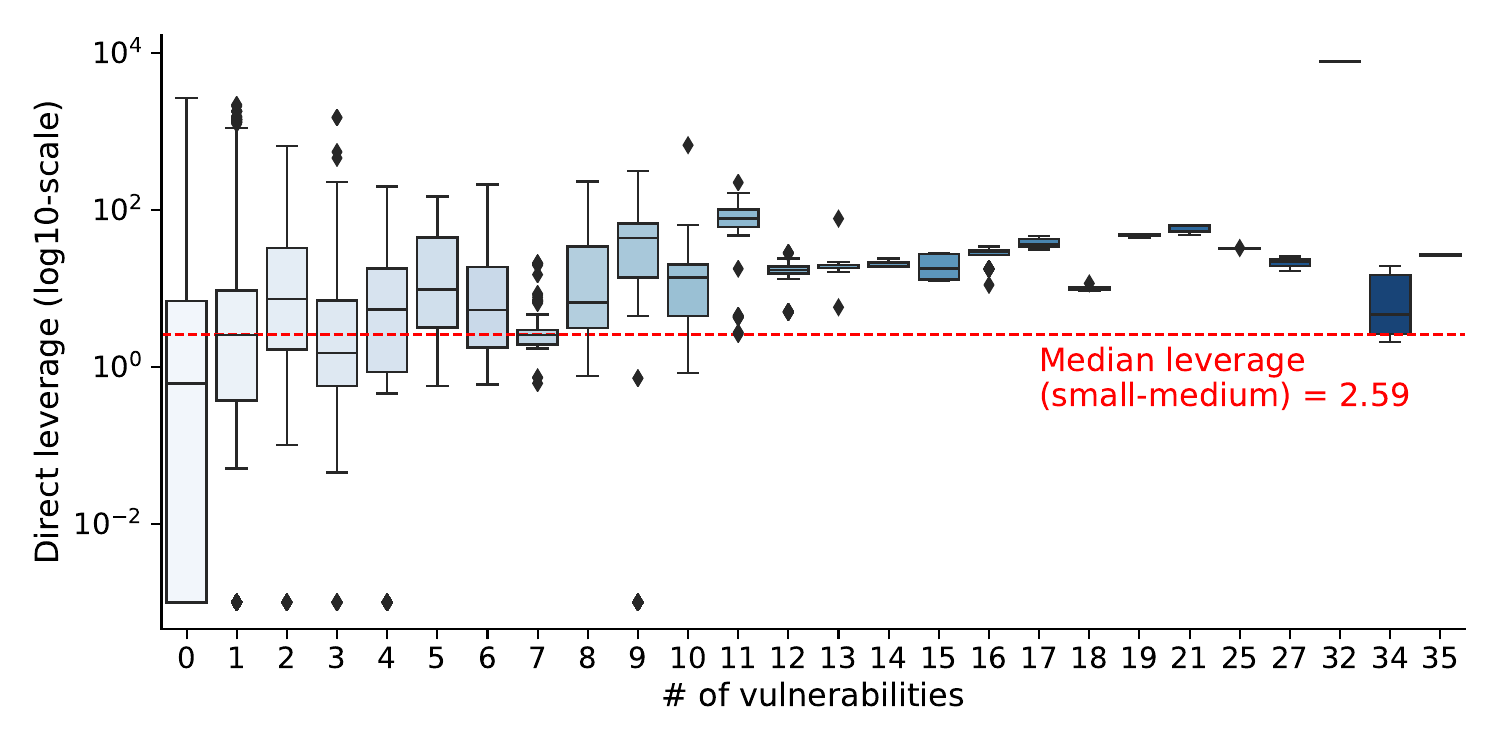}
\caption{Direct technical leverage per library vs Number of vulnerabilities for small-medium library releases.}
\label{vul_max_small}
\end{figure}

\begin{figure}[tb]
\centering
\includegraphics[width=.9\linewidth]{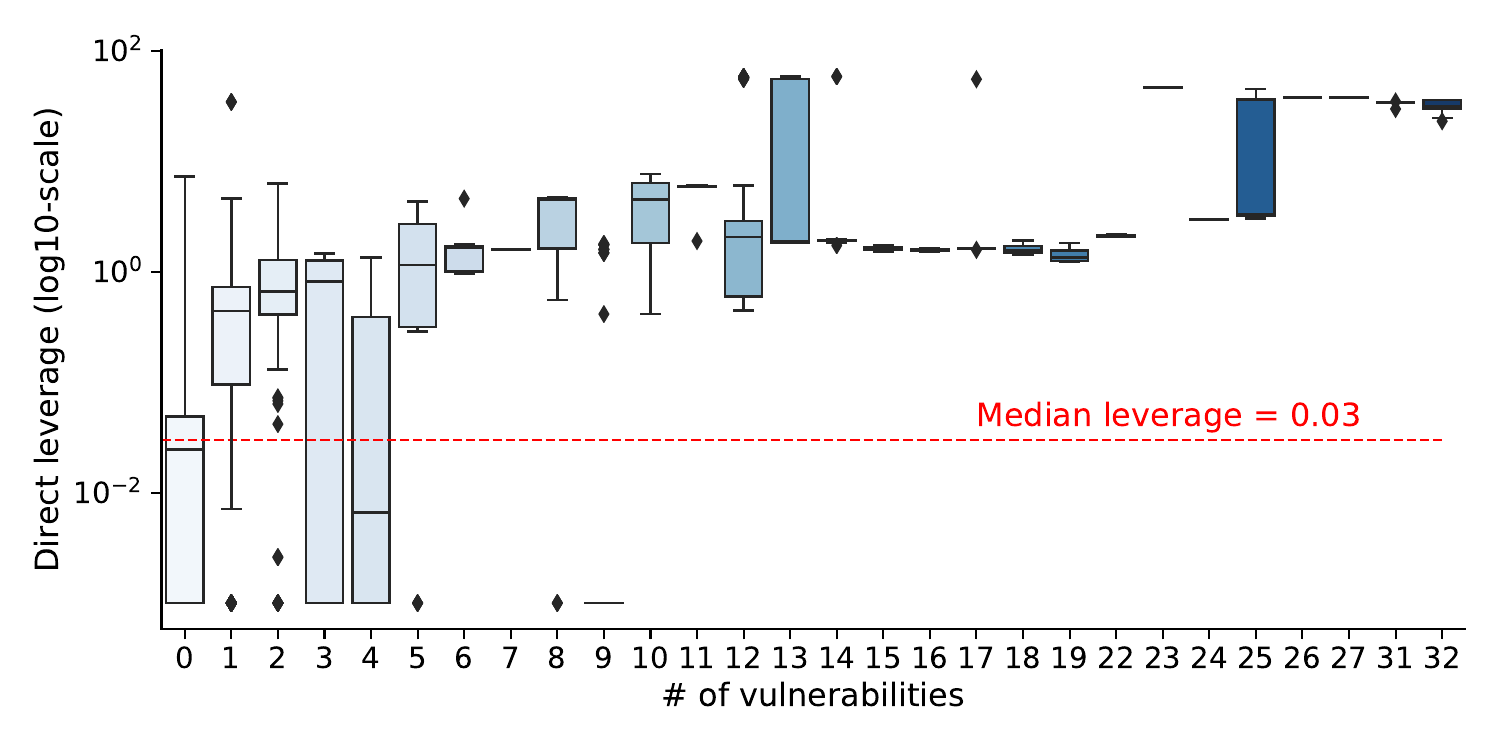}
\caption{Direct technical leverage per library vs Number of vulnerabilities for large library releases.}
\label{vul_max_large}
\end{figure}

To better visualize the relationship between the direct technical leverage and the number of reported vulnerabilities, we present this relation in Figure~\ref{vul_max_small} for small-medium libraries and Figure~\ref{vul_max_large} for large libraries.
In both figures, we present the frequency of library releases grouped by the number of vulnerabilities (x-axis) and their respective technical leverage (y-axis).  
From Figure \ref{vul_max_small}, we observe that small-medium libraries ($<$10KLOC) with direct technical leverage higher than 2.5 (median-red line), have higher chances of facing more vulnerabilities. 
Looking at the number of vulnerabilities with the range [5-35] in Figure \ref{vul_max_small}, one can observe that numerous library releases have direct technical leverage higher than 2.5. 
This difference is considerably more prominent in large libraries, as shown in Figure~\ref{vul_max_large}.
Almost all releases with 5 or more vulnerabilities have a high direct technical leverage (above the red line). In contrast, libraries that have 4 or fewer vulnerabilities have low technical leverage.
This confirms the intuition that the direct technical leverage metric is a good indicator of the degree of risk (vulnerability exposure) to which the library is exposed.

\begin{figure}
\centering
\includegraphics[width=.8\linewidth]{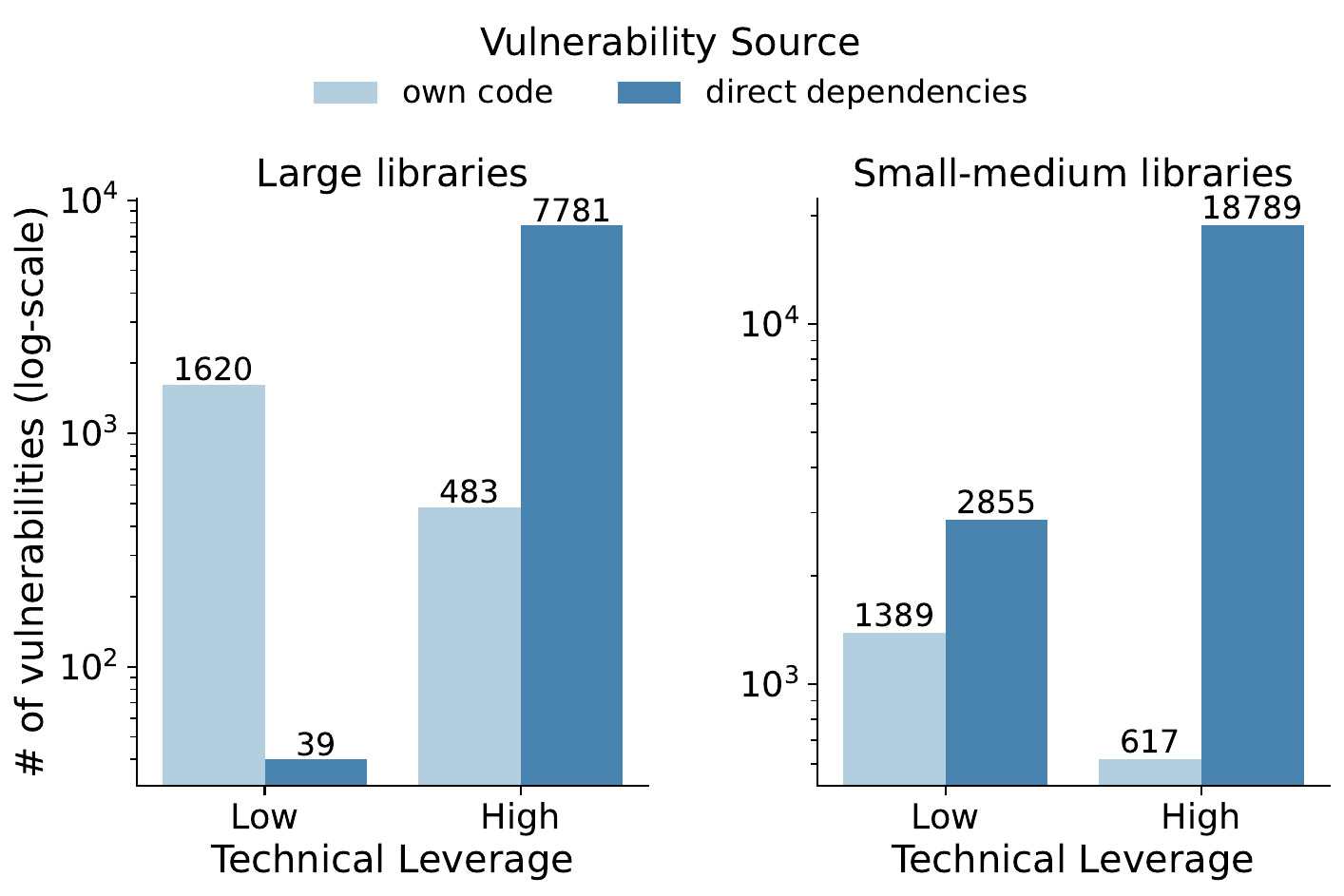}
\caption{Count of all vulnerabilities reported in small-medium and large library releases per vulnerability source and technical leverage. Note that the y axis is in log-scale.}
\label{source_vul}
\end{figure}

So far, we have shown that direct technical leverage has an impact on the number of vulnerabilities across small-medium and large library releases. 
To better understand the source of vulnerability, we compare the number of vulnerabilities found in their own code versus in the dependencies as shown in Figure~\ref{source_vul}.
In libraries with low technical leverage, we find that the main source of vulnerabilities is their own code, in large libraries (1,620), whereas, direct dependencies are the main source of vulnerabilities for small-medium size libraries (2,855). 
On the other hand, dependencies become the primary source of vulnerabilities in libraries with high technical leverage.
Large libraries reported 7,781 vulnerabilities from dependencies against 483 vulnerabilities in their own code. 
While the small-medium libraries dependencies reported a total of 18,789 vulnerabilities versus 617 reported in their own code. 
This shows that high direct technical leverage comes at a cost: developers are more likely to have vulnerability reported in their dependencies than in their own code, for both small-medium and large libraries in the NPM ecosystem.

\textbf{Comparison to the Maven ecosystem.} 
In both Maven and NPM, shipping more times your code base will increase the risk of being exposed to vulnerabilities. 
However, the risks in NPM far surpass the ones reported by the original study in Maven.
In Maven, projects are 60\% more likely to become vulnerable when they rely on high technical leverage \cite{massacci2021technical}. 
However, in NPM, high technical leverage increases the risk of vulnerabilities by 4 times for small-medium libraries and 7 times for large libraries. 
\begin{tcolorbox}
 \textbf{Key Takeaway:} Direct technical leverage is associated with high security risks in both NPM and Maven ecosystems. In NPM, small to medium-sized libraries using high direct technical leverage face a fourfold increase in reported vulnerabilities, while large libraries with high technical leverage have a 7.6 times higher risk of being affected by reported vulnerabilities. This observation was also presented in the Maven ecosystem, where small to medium-sized libraries exhibit a 60\% increase in the risk of releasing a vulnerable version if they leverage more code from dependencies. 
\end{tcolorbox}


\subsection{RQ4: To what extent do the findings about the technical leverage observed for direct dependencies hold for level-1 transitive dependencies?}
\textbf{Motivation:} So far, our study has focused on assessing the opportunities and risks related to direct dependencies in NPM to make our results comparable to the replicated study~\cite{massacci2021technical}, modelled through the lens of direct technical leverage. 
Direct dependencies are only the first level of dependencies of a project, and often represent a small share of the total dependencies. 
As libraries also reuse other libraries' code, transitive dependencies quickly become the vast majority of dependencies in a project~\cite{latendresse2022not,prana2021out,decan2017empirical}. 
For example, Decan et al.~\cite{decan2017empirical} analyzed 300K NPM packages and found half of the analyzed packages have at least 22 transitive dependencies, and a quarter have at least 95 transitive dependencies. These packages are often deployed in JavaScript applications in production~\cite{latendresse2022not}, where the risks of vulnerabilities are potentially harmful.

\noindent
Among various ecosystems, NPM stands out for its higher code reuse compared to the Maven ecosystem. Illustratively, data from the "State of the Developer Ecosystem 2022" survey highlights an evident example: the NPM ecosystem showcases a greater prevalence of code reuse in comparison to Maven. This survey, drawing insights from responses provided by over 29,000 developers worldwide, underscores the inclination of developers within the NPM community to actively leverage existing packages and libraries \cite{devecosystem}. This robust culture of code reuse, however, introduces a notable risk factor when considering transitive dependencies. Previous research in \cite{dusing2022analyzing} has demonstrated that transitive dependencies serve as a potential vector for vulnerabilities, with the ability to impact libraries through extensive chains of dependencies. Another study in \cite{lauinger2018thou} conducted a comprehensive examination of JavaScript open source projects. Their study highlighted a noteworthy discovery: transitive dependencies within a project are more prone to vulnerabilities. In addition, developers often leverage existing dependencies to expedite the development of their own libraries.

\noindent
The impact of transitive dependencies on project quality motivated us to extend our replication study to include a preliminary assessment of the opportunities and risks of technical leverage, also considering transitive dependencies.

\noindent
\textbf{Approach.} We expanded our definition of direct technical leverage also to include the first level of transitive dependencies ($\lambda_{dir + trans1}$), as shown in Equation 5.
The first level of transitive dependency is the direct dependencies of a project's direct dependencies, i.e., dependencies with depth 2 in a dependency graph. 
We choose to include only the first level of transitive dependencies to keep our study feasible. 
Including all transitive dependencies, while ideal to provide a complete picture of our analysis, is too computationally expensive as their number grows exponentially~\cite{decan2017empirical}.
More importantly, prior work shows that only considering the direct dependencies and the first level of transitive dependencies already accounts for the vast majority of vulnerabilities in npm packages~\cite{mir2023effect}.
We collect the first level of transitive dependencies for all library releases (14,042 releases) in our dataset described in Section~\ref{data_selection} and use the same experimental settings described in Section~\ref{data_selection} to analyze the first level of transitive dependencies for all releases in our dataset. In particular, we analyzed 320,253 dependencies to understand the impact of including transitive dependencies on the difference in technical leverage between small-medium and large libraries, the time interval between library releases, and the risk of including more vulnerabilities.
\\

\vspace{.2cm}
 \vspace{1.1\baselineskip}
 \begin{equation} \label{eqs}
    \lambda\textsubscript{dir + trans1} = 
    \frac{
        \tikzmarknode{dir}{\highlight{red}{L\textsubscript{dir}}} + 
        \tikzmarknode{tran}{\highlight{OliveGreen}{L\textsubscript{trans1}}} 
        }
        {\tikzmarknode{own}{\highlight{blue}{L\textsubscript{own}}}
    } 
\end{equation}

\begin{tikzpicture}[overlay,remember picture,>=stealth,nodes={align=left,inner ysep=1pt},<-]
    \path (dir.north) ++ (0,1em) node[anchor=south east,color=red!67] (dirlabel){size of direct dependencies};
    \draw [color=red!87](dir.north) |- ([xshift=-0.3ex,color=red]dirlabel.south west);
    \path (own.south) ++ (0,-1em) node[anchor=north west,color=blue!67] (ownlabel){size of own code};
    \draw [color=blue!57](own.south) |- ([xshift=-0.3ex,color=blue]ownlabel.south east);
    \path (tran.north) ++ (0,1em) node[anchor=south west,color=OliveGreen!67] (tranlabel){size of first level of transitive deps};
    \draw [color=OliveGreen!87](tran.north) |- ([xshift=-0.3ex,color=green]tranlabel.south east);

    
\end{tikzpicture}
\vspace{1.1\baselineskip}

\vspace{.2cm}

\begin{table}[]
    \centering
    \caption{Comparison of results from direct technical leverage ($\lambda_{dir}$) and including the first level of transitive dependencies ($\lambda_{dir + trans1}$).}
    \label{tab:transitive_dependencies_results}
    

    


\begin{tabular}{p{4.5cm}|p{1.2cm}p{1.2cm}|p{1.1cm}p{1.1cm}}
\toprule
& \multicolumn{2}{c|}{\textbf{Small-medium libraries}} & \multicolumn{2}{c}{\textbf{Large libraries}} \\ 
\midrule
& $\lambda_{dir}$ & $\lambda_{dir + trans1}$ 
& $\lambda_{dir}$ & $\lambda_{dir + trans1}$  \\ 
\midrule
Median Technical Leverage (RQ1) & 2.5 & 7.7 & 0.03 & 0.06  \\ 

\midrule

Impact of technical leverage on release cycle (RQ2)

& Small positive impact    & Small positive impact & -- & -- \\ 

\midrule

Vulnerability risks for libraries with high technical leverage (RQ3) & 4x & 6.7x & 7.6x & 21x \\ 
\bottomrule
\end{tabular}%

    

\end{table}
\noindent
In this study, we consider both direct dependencies and transitive dependencies for the vulnerability impact analysis. By conducting separate analyses, we aim to gain a comprehensive understanding of the overall upper bound impact of vulnerabilities within the software ecosystem. To this aim, we first resolve the dependencies of each library release, using the available dependency versions at the release time (same method described in Section 3.2). Then, for each resolved dependency of a library release, we search the Snyk dataset to determine if the dependency is affected by vulnerabilities. Each vulnerability reported in Snyk has a range of affected versions, and we use this affected version range to infer whether dependencies are considered vulnerable.  This information is aggregated back to the library release, by counting vulnerable dependencies and assessing their distribution over the severity levels.  By adhering to this criterion, we aim to ensure a thorough and accurate evaluation of the vulnerability status for each library release in our dataset. The same process is applied for transitive dependencies. To elaborate more,  in evaluating technical leverage, encompassing level 1 transitive dependencies, we assessed the quantity of vulnerable transitive dependencies. Consequently, we categorized the library release as vulnerable if the cumulative vulnerability count from both direct and level-1 transitive dependencies reaches at least one.

\noindent
\textbf{Result.} We present an overview of the results that include the first level of transitive dependencies in Table~\ref{tab:transitive_dependencies_results}.
Overall, the observations we presented through RQ1-RQ3 hold. 
Technical leverage presents opportunities to ship more code (RQ1) and releases faster in the case of small-medium libraries (RQ2), but comes at the cost of a higher risk of vulnerabilities (RQ3).
Expectedly, once we include the first level of transitive dependencies, the magnitudes of opportunities and risks change: we can observe a significant increase in code leveraged - as transitive dependencies add to the direct technical leverage - and observe a much higher risk of vulnerabilities (see the comparisons between $\lambda_{dir}$ and $\lambda_{dir + trans1}$ in Table~\ref{tab:transitive_dependencies_results}).  
In the following, we discuss the differences and implications of this expanded experiment. 
\\

\begin{figure*}
\includegraphics[width=\linewidth]{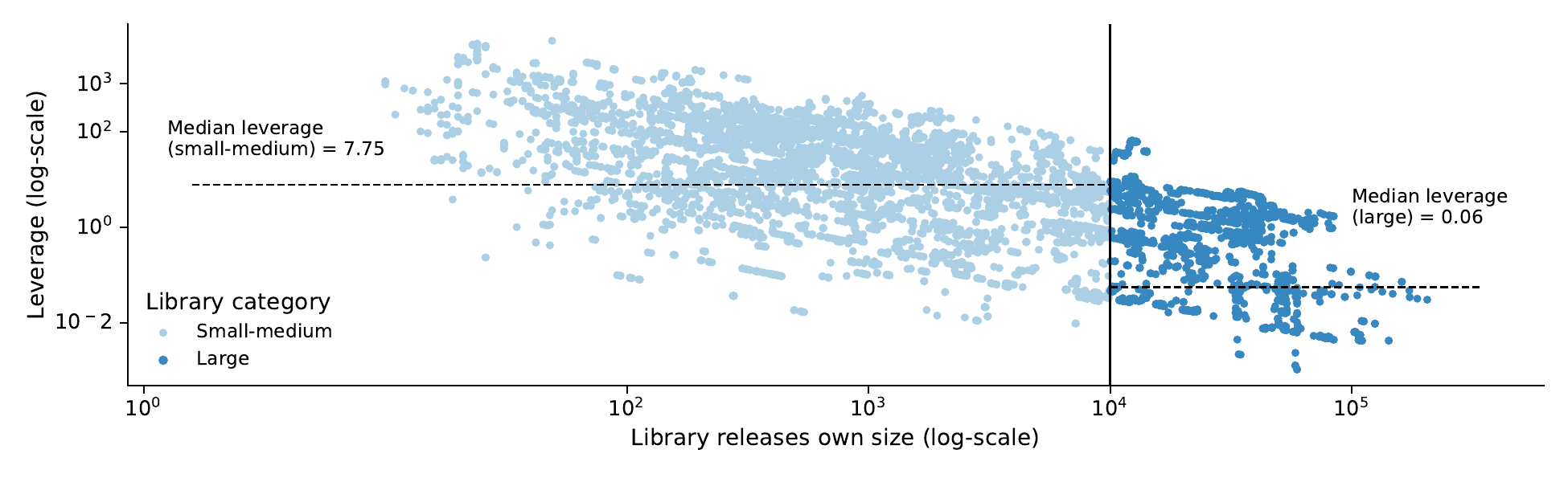}
\caption{The technical leverage ($\lambda_{dir + trans1}$) of 14,042 library releases per library size in LOC. }
\label{trans_lev}
\end{figure*}

\noindent
\textbf{Technical leverage.}
As Table~\ref{tab:transitive_dependencies_results} shows, including transitive dependencies increased the amount of code leveraged by libraries in FOSS by 2-3 times, compared to just the direct technical leverage.
Small-medium sized libraries leverage 7.7 times their code in FOSS considering $\lambda_{dir + trans1}$ and large libraries saw their amount of code borrowed double, from 3\% to 6\%.
Figure~\ref{trans_lev} presents the relation between the own code size of the libraries and technical leverage when including transitive dependencies. Compared to the results shown in Figure~\ref{direct_lev}, the overall picture remains: small-medium libraries proportionally ship other people’s code more than large libraries. 
\\


\noindent
\textbf{Time interval between library releases.}
In RQ2, we report that small-medium libraries that leverage more FOSS code have shorter release cycles, albeit with a small effect size, as shown in Table~\ref{tab:transitive_dependencies_results}. 
To understand the impact of including transitive dependencies on the release speed of both small-medium and large libraries, we built a multivariate linear regression model to capture the impact of technical leverage on the time interval between library releases (similar to RQ2). 
Similar to the results in RQ2, we observe that the coefficient of the technical leverage ($log(\lambda_{dir+trans1})$) has a significant negative coefficient (see Table~\ref{table_reg_L1}), indicating that the technical leverage ($\lambda_{dir+trans1}$) is associated with a small effect of shortening release cycles. 
To understand the practical effect of technical leverage in the release cycle, we performed the same experiment as described in RQ2, by  simulating the effect on the release cycle, if we increase all libraries technical leverage by 4 times. 
The result of this experiment indicated that the impact of 4 times increase in ($\lambda_{dir+trans1}$) on release cycle is similar to considering direct dependencies (1 day).


\begin{table*}[h]
    \centering
    
    \caption{Linear regression model fit to check the correlation between \(\lambda\textsubscript{dir+trans1}\), \(\theta\), \(\rho\), and the release time interval.}
    \begin{adjustbox}{max width=\columnwidth}
\begin{tabular}{ll|ll|ll}
        \toprule
        \textbf{Description} &
        \textbf{Coefficients} & 
        \multicolumn{2}{c}{\textbf{Small-medium libraries}} &
        \multicolumn{2}{|c}{\textbf{Large libraries}}\\
         &
         & 
         \textbf{Estimate} &
         \textbf{P-value} &
         \textbf{Estimate} &
         \textbf{P-value} \\

         \midrule
         
         Intercept& 
         1& 
         0.4406&
         0.000 &
         0.3984 &
         0.000 
          \\
         
         Technical Leverage &
         log(\(\lambda\textsubscript{dir+tarns1} \)) &
         -0.0349   &
         0.000 &
         \textbf{0.0040} &
         \textbf{0.688 } 
        \\
         
         Change distance &
         log(\(\rho\)) &
          0.1041 &
         0.000 &
         0.0792 &
         0.000
         \\
         
         Change in total code  &
         cos(\(\theta\) - 45\textsuperscript{o}) &
         0.1800 &
         0.000 &
         \textbf {0.0435}  & 
         \textbf {0.132}
         \\

         Change in own code&
         sin(\(\theta\)) &
         \textbf {0.0055}  &
         \textbf {0.695} &
         \textbf {0.0328}   &
         \textbf {0.154}
         \\

        Previous release interval &
        $log(\Delta_{x-1} + 1)$&
          0.2540  &
         0.000&
         0.2213     &
         0.000
\end{tabular}

\end{adjustbox}
\footnotesize

\begin{flushleft}

For small-medium libraries: Root mean square error(rmse) =0.65, R-squared( R\textsuperscript{2}) = 0.160, Adj. R-squared ($\overline{R}$\textsuperscript{2}) = 0.160\\
For large libraries: Root mean square error(rmse) =0.56, R-squared( R\textsuperscript{2}) = 0.118, Adj. R-squared  ($\overline{R}$\textsuperscript{2}) = 0.117

\end{flushleft}

\label{table_reg_L1}
\end{table*}

Similarly to the results reported in RQ2, large libraries seem not to benefit from high technical leverage, neither direct nor when we include the first level of transitive dependencies. 
\\

\begin{figure}
\centering
\includegraphics[width=.7\linewidth]{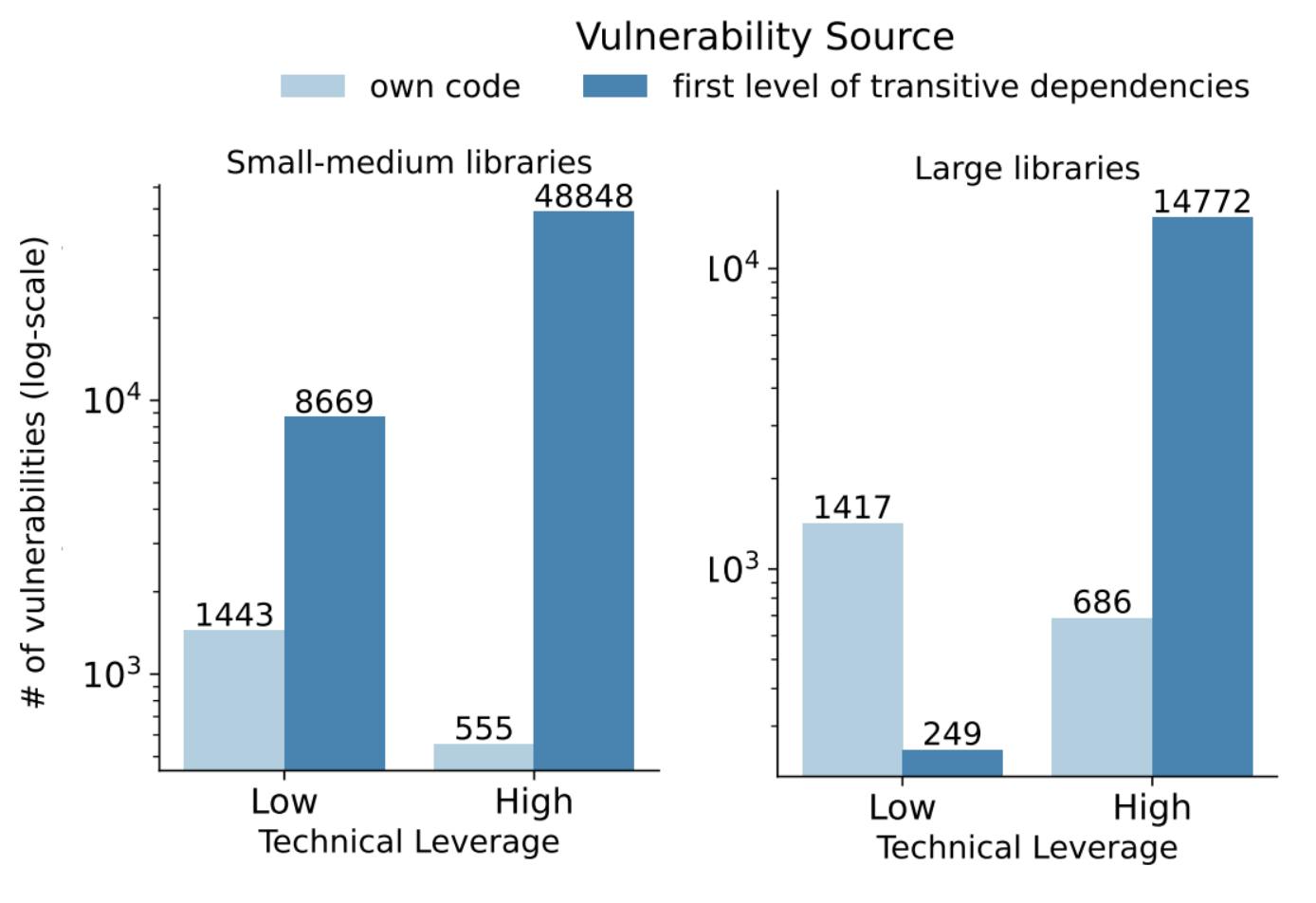}
\caption{Count of all vulnerabilities reported in small-medium and large library releases per vulnerability source and technical leverage. Note that the y-axis is in the log scale.}
\label{source_vul_trans}
\end{figure}

\noindent
\textbf{Risk of including more vulnerabilities.}
We use the same method presented in RQ3 to assess the risks of including vulnerabilities that originate in both the direct and the first level of transitive dependencies. 
Figure~\ref{source_vul_trans} presents the frequency of library releases grouped by the number of vulnerabilities (x-axis) and their respective technical leverage (y-axis) for small-medium and large library releases. 
The primary source of vulnerabilities for both small-medium and large library releases remains their dependencies. 
Our results show that technical leverage increases the magnitude of security risks for both small-medium and large libraries, compared to direct dependencies discussed in RQ3 (see Table~\ref{tab:transitive_dependencies_results}).
Small-medium libraries that rely on high technical leverage have their vulnerability risk increase from 4 to 6.7 times when we include the first level of transitive dependencies in the analysis. 
The risk in large libraries increased even further.
Large libraries with high technical leverage have their risk increase from 7.6 times to 21 times when we consider the first level of transitive dependencies. 

\noindent
It is important to highlight, however, that when we consider transitive dependencies, the risk of overcounting vulnerabilities also increases.
We opted to include transitive dependencies to provide yet another frame of reference for the analysis of opportunities and risks of using open-source software. Those studies have discussed that transitive dependencies inflate the problem of vulnerabilities, saying that a sizeable number of vulnerabilities do not represent a threat to software security because the vulnerable part of the library is never used in a software project. A recent investigation by Mir et al. \cite{mir2023effect} in the Maven ecosystem revealed that less than 1\% of packages demonstrate a reachable call path to vulnerable code within their dependencies. This percentage is notably lower when contrasted with the outcomes derived from a simplistic dependency-based analysis. Decan et al. \cite{decan2018impact} examined the influence of security vulnerabilities on the NPM dependency network. Their research revealed that about 15\% of vulnerabilities are categorized as high risk, as they are addressed subsequent to their publication date. Zimmermann et al. \cite{zimmermann2019small} examined security threats within the NPM ecosystem and uncovered a notable finding: a relatively small number of JavaScript packages have the potential to affect a significant portion of the NPM ecosystem. This suggests that if maintainer accounts were compromised, they could be leveraged to inject malicious code into a majority of NPM packages. In a recent study, Liu et al.  \cite {liu2022demystifying} investigated vulnerability propagation and its evolution within the NPM ecosystem. By constructing a comprehensive dependency knowledge graph, they uncovered notable findings, including the discovery that 30\% of package versions are impacted by overlooking vulnerabilities in direct dependencies. Another prior study by Zapata et al. \cite{zapata2018towards} checked 133k websites and found 37\% websites use at least one JavaScript library with a known vulnerability. Furthermore, they found that libraries included transitively are more likely to be vulnerable, which is aligned with our reported findings.

\begin{tcolorbox}
\textbf{Key Takeaway:} Results show similar opportunities and risks, compared to considering only
direct dependencies, but with a significant change in the magnitude of both
benefits and downsides.
\end{tcolorbox}

\section{Implications}\label{impl}

In this section, we highlight the most important implications of the results of our empirical study for both developers and researchers.\\

\noindent
\textbf{Small libraries size can be deceiving if not accounted for the technical leverage.}  
Previous studies have shown that practitioners consider the size of the libraries when selecting a library for their project~\cite{abdalkareem2017developers} as a small code base implies easier project maintenance. 
Our results in RQ1, however, show that small libraries in NPM tend to leverage 2.5 times their own code as direct library dependencies, indicating that a larger part of the code-base of a small library actually belongs to their dependencies' code base.
Currently, this information is hidden from developers, who can only estimate the technical leverage by accounting for the number of dependencies. 
The research community should work towards making the information about technical leverage more visible for practitioners, that can then make a better informed decision when selecting a library for their project.
\\

\noindent
\textbf{Technical leverage tends to only increase over time.}
Our results in RQ1 show that dependency code tends to stay the same or increase over time in the majority of the releases. 
Maintainers only rarely make the effort of reducing the dependency code over time, depicted as change direction ($\theta$ = 180$\degree$), and this can lead to significant bloating in the software project.
The bloat caused by maintaining unnecessary dependencies has strong consequences to the software quality, as it increases the attack surface for security vulnerabilities~\cite{pashakhanloo2022pacjam,soto2021comprehensive,koishybayev2020mininode,azad2019less} and it impacts the application's performance~ \cite{soto2021comprehensive,azad2019less,quach2019bloat}, overall increasing the project maintenance costs~\cite{soto2021comprehensive,soto2022coverage,soto2021longitudinal}.  
To proactively address and mitigate this issue, we recommend the implementation of a structured process for developers to systematically remove or update dependencies that are no longer essential or possess viable alternatives.
\\

\noindent
\textbf{Benefits and risks of technical leverage are magnified in the NPM, compared to Maven libraries.}
In both ours and the replicated study~\cite{massacci2021technical}, results point to the double-edged sword characteristic of technical leverage: speeding up software releases but incurring a higher risk of exposure to public vulnerabilities.
However, both benefits and risks seem to be magnified in NPM. 
On the good side, higher technical leverage seems to yield faster releases at a much higher rate in NPM than in Maven: leveraging 4 times one's code is associated with faster release in NPM, but incurs a few days delay in Maven libraries.
On the bad side, libraries with high technical leverage in NPM have 4-7 times more chances of being affected by vulnerabilities, while in Maven, the results point to a 60\% increase of vulnerability exposure. 
Further investigation is necessary to establish the cause of these vast differences across ecosystems. However, practitioners should take these results as a motivation to establish practices of open source governance, such as, including an explicit process for vetting new libraries, control over the bill of materials (BOM), and DevSecOps practices~\cite{Stan:2022:OpenSource}.
To further elaborate on these practices, a set of actionable examples is introduced. For the practice of establishing an explicit process for vetting new libraries, developers may create a comprehensive checklist for evaluating the suitability of potential libraries before incorporating them into the project. This checklist could encompass critical factors such as community support, security considerations, licensing, and alignment with project goals.
To further detail the practice of controlling the Bill of Materials (BOM), developers may implement a centralized dependency management system designed to maintain a thorough and up-to-date BOM. This system plays a crucial role in providing real-time insights into dependencies, their respective versions, and any associated security vulnerabilities. By doing so, it empowers proactive decision-making within the development process. Finally, for the DevSecOps Practices, developers may Integrate security checks seamlessly into the continuous integration/continuous deployment (CI/CD) pipeline. This ensures that security considerations are addressed at every stage of development, reducing the likelihood of introducing vulnerabilities during the software development life cycle.
\\

\noindent
\textbf{We need more fine-grained methods to assess the risks and benefits of technical leverage.}
Both the replicated study and ours are based on the direct technical leverage, calculating benefits and risks considering only the direct dependencies. 
The (complete) technical leverage considers both direct and transitive dependencies. 
However, the lack of more fine-grained methods for dependency usage adds numerous challenges to a study that aims to investigate technical leverage.
Without fine-grained methods that analyze the project's source code and infer what modules are imported, computing the technical leverage will lead to substantially overestimated results~\cite{zapata2018towards}.
Previous studies have shown that the majority of transitive dependencies are not actually leveraged by software projects and can be safely removed~\cite{Ponta:21:Bloated}. 
To expand the study, on a large scale, from direct technical leverage to technical leverage, we need to further the development of fine-grained methods and datasets~\cite{Paolo:21:Finegrained}.
Such methods will help practitioners in better assessing dependencies that are relevant for security and maintenance, as well as help researchers study the pros and cons of technical leverage. 
Thus, instead of solely relying on broad metrics, such as the frequency of code reuse, developers could  consider more granular measurement metrics. For instance, analyze the impact of technical leverage on specific software components, modules, or functions. This fine-grained approach can unveil insights into the effectiveness and risks associated with different levels of code reuse. Moreover, it is possible to explore the temporal aspects of technical leverage by examining how the effectiveness and risks evolve over time. This could involve tracking changes in code reuse patterns, identifying long-term benefits or downsides, and understanding the sustainability of technical leverage strategies.
\\



\noindent
\textbf{Technical leverage's opportunities and risks vary significantly across software ecosystems.}
The purpose of our study was to contrast how technical leverage is used across two major software ecosystems: NPM and Maven. 
Our findings show that the opportunities/risks of technical leverage vary from the replicated study. Hence, our study motivates the need for more research to examine the benefits/downsides of code reuse in other ecosystems (e.g. PyPI, Go ecosystem). By extending our focus beyond the confines of NPM and Maven, researchers can gain a more comprehensive understanding of the implications of technical leverage on software development practices. Thus, employing technical leverage empowers developers to discern the evolution and security posture of a software ecosystem. This insight becomes pivotal for developers as they evaluate the viability of adopting, updating, or refraining from changes to a library. By leveraging technical insights, developers can make informed decisions, enhancing their ability to navigate and contribute to the ever-evolving landscape of software development.
We publish a replication package, including our entire methodology and scripts to help foment new studies on the topic. 
\\

\section{Related Work} 
\label{related_work}


Numerous studies have explored the idea of comparing similar aspects across software
ecosystems~\cite{Bogart:21:PracticesPolicies,decan2016topology,decan2019empirical}. 
Decan et al.~\cite{decan2019empirical} empirically compared the evolution of dependency network in seven ecosystems. They found that there are some differences across ecosystems due to the differences in ecosystems' policies and community's practices. 
Bogart et al.~\cite{Bogart:21:PracticesPolicies} compared breaking changes practices across 18 software ecosystems, reporting numerous differences in how each ecosystem community handles breaking changes. Breaking changes in software packages refer to significant modifications that can disrupt existing functionality in dependent applications, requiring developers to adjust their code for compatibility \cite{raemaekers2014semantic,sawant2018understanding}. These changes encompass API modifications, dependency updates impacting compatibility, removal or deprecation of features. Managing these changes is crucial for maintaining application stability and functionality.
Abdalkareem et al. \cite{abdalkareem2020impact} investigated the NPM and PyPI ecosystems. The study showed that up to 16\% of Python and JavaScript packages are trivially small (i.e., have less than 250 lines of code), but are highly depended upon in those ecosystems. 
Studies that establish parallels across ecosystems are crucial for our understanding on the dynamics of software development. 
The aforementioned works motivate us to complement the replicated study~\cite{massacci2021technical} by investigating technical leverage and its associated effects on another major ecosystem, the NPM. 
\\

\textbf{Security Vulnerabilities \& Dependencies.} 
Given their importance, many works have focused on better understanding the dynamics involving security vulnerabilities in the software ecosystem~\cite{pham2010detection,hejderup2015dependencies}.
Pham et al. \cite{pham2010detection} conducted an empirical study on thousands of vulnerabilities, finding that most vulnerabilities are recurring due to software code reuse. 
Zerouali et al. \cite{zerouali2021impact} conducted a comparison between RubyGems and NPM ecosystem to quantify the impact of security vulnerabilities. 
The study found that for both ecosystems, the time required to discover vulnerabilities is increasing, and the number of vulnerabilities is increasing faster. However, the results showed that the vulnerabilities are discovered faster in NPM compared to RubyGems.
Alfadel et al.~\cite{Alfadel:2021:Python} compared the vulnerability reporting process of PiPy and NPM, reporting that several differences in the ecosystem's policy has an impact on the exposure of package's vulnerabilities.

Nappa et al. \cite{nappa2015attack} identified several security threats caused by shared libraries distributed among ten popular client applications on Windows.
Gkortzis et al. \cite{gkortzis2019double} empirically investigated 301 open source java projects to explore the relationship between software reuse and security vulnerabilities, based on static analysis of the source code. They found that the amount of potential vulnerabilities in both native and reused code increases with larger project sizes. 
Moreover, Gkortzis et al. \cite{gkortzis2021software} empirically investigated 1,244 open-source Java projects to further explore the relationship between software reuse and security vulnerabilities. They found that large projects are highly associated with a higher number of vulnerabilities, and  the number of dependencies is considerably correlated to its number of vulnerabilities. 

Pashchenko et al. \cite{pashchenko2018vulnerable} studied how much code reuse in the SAP ecosystem is affected by vulnerabilities, they analyzed the top 200 open source Maven projects that were reused in SAP. The authors found that 13\% of the direct and transitive dependencies were affected by at least one vulnerability. 
Cox et al. \cite{cox2015measuring} analyzed 75 Java projects managed by Maven. They found that projects using outdated dependencies were 4 times more likely to have security issues compared to the up-to-date dependencies.
Moreover, many studies illustrated that free open source software dependencies are widely used by commercial projects as well as FOSS projects, but they lack the proper maintenance \cite{lauinger2018thou,cox2015measuring}. About 81.5\% of the studied systems remain with outdated dependencies \cite{decan2018impact,pashchenko2018vulnerable}. Such behaviour introduces serious bugs and security vulnerabilities \cite{kula2018developers}. 

As a replication study, our study is the most closest to the study by Massacci and Pashchenko~\cite{massacci2021technical}.
The differences between the Maven and NPM ecosystems motivated us to replicate the \cite{massacci2021technical} on the NPM ecosystem. 
We believe that our study complements the previous study by quantifying the impact of the newly introduced metrics \cite{massacci2021technical} on the release speed and health of the NPM libraries for both direct and level-1 transitive dependencies. 
Moreover, our study highlights some differences between Maven and NPM ecosystems which require further investigation such as NPM libraries have considerably lower direct technical leverage compared to Maven libraries.


\section{Threats to Validity}
\label{threat}
In this section, we discuss the threats to the internal, construct, and external validity of our study.
\noindent
\subsection{Internal Validity}
Concerns confounding factors that could have influenced our results.
In evaluating the vulnerability impact analysis, an internal threat to validity concerns overcounting vulnerable dependencies. This threat is particularly problematic when it comes to the inclusion of transitive dependencies. The risk of overcounting vulnerabilities associated with transitive dependencies has been recognized in previous research \cite{decan2018impact,liu2022demystifying}, emphasizing that a substantial portion of these vulnerabilities may not pose an actual threat to software security.  Given that we perform a large-scale analysis in the npm ecosystem, there are no JavaScript methods for precisely identifying if a vulnerable portion of a transitive dependency poses a threat to its downstream dependents. Thus, our findings should be interpreted as an upper bound for the risks of high technical leverage, i.e., applications are likely to experience a lower risk in practice.
The diversity of tasks within npm packages spanning data science, networking, and web development poses a potential threat to internal validity. This stems from the potential influence of specific tasks on the technical leverage of individual libraries. For instance, a data science package, with features like machine learning models and numerical computations, may exhibit larger size and significantly higher technical leverage compared to a package designed for web development. Failure to appropriately address these task-specific influences may compromise the internal validity of the study. However, it's crucial to note that our study focuses on an abstract view of the NPM ecosystem, comparing it to Maven.
Another potential concern lies in vulnerability analysis, which involves identifying vulnerabilities from the Snyk database. To alleviate this threat, we selected 50 samples and then initiated a POST request to the Open Source Vulnerabilities (OSV) API endpoint at https://api.osv.dev/v1/query. This request contained pertinent data such as the package name and version. Our findings validated the robustness of our results.

\noindent
\subsection{Construct Validity} Considers the relationship between theory and observation, in case the measured variables do not measure the actual factors. 
In our study, we focus on including the most popular NPM packages by selecting the top-most depended upon packages.  
Numerous other criteria are used to assess popularity, e.g., the number of stargazers and number of downloads, which could result in a different set of studied libraries. 
We still believe that our dataset is fairly representative of the most popular NPM, including libraries such as react, Lodash, angular, and Vue.   
Another threat is related to the construct of small-medium versus large libraries. 
We choose the threshold of 10KLOC to distinguish small-medium size libraries from large size libraries. 
Selecting different thresholds might lead to different results. 
To alleviate this threat, we explored different thresholds (i.e., 5, 10, 15, and 20 KLOC), and we found that the final findings persist across the different thresholds.
The choice of vulnerabilities dataset used in our analysis poses a potential threat that may have impacted our findings. 
We rely on the Snyk.io dataset as the only source for vulnerabilities. 
Our choice of Snyk is inspired by prior work \cite{decan2018impact,massacci2021technical,chinthanet2021lags}, and the Snyk team continuously monitors widely used libraries and their associated vulnerabilities \cite{snyk}. 
So, we expect our vulnerabilities dataset to be reasonably accurate and not impact our results in a significant manner. 
In our analysis, we utilized Ordinary Least Squares (OLS) regression, well-suited for continuous dependent variables. However, a potential concern arises when independent variables possess different scales, potentially impacting the regression model's coefficient interpretation. To mitigate this threat, we thoroughly examined the ranges of all independent variables and confirmed that they fall within reasonable bounds. Another potential concern arises from the incorporation of a log scale in the regression model. However, we justify this choice by illustrating the distribution of each variable. The decision to apply the log function is motivated by the observation of long tails in the distributions, highlighting the need for a transformation to address skewed data and improve the model's robustness. 

\noindent
\subsection{External Validity} Concerns the generalization of our results to other software ecosystems.
While some of our findings are inline with the results of the replicated study on the Maven ecosystem, other findings
are not.
It is expected that our results are representative of the most popular libraries in the NPM ecosystem but may not generalize to other ecosystems (e.g., PiPy, CRAN). 
Furthermore, our study focuses on analyzing libraries in NPM, which have specific characteristics and may differ from other types of projects (e.g., applications).
The methodology we replicate in the study could (and should) be applied to other software ecosystems to establish a more comprehensive view of the impact of direct technical leverage on ecosystems.
Another important consideration for the external validity of our study is the potential limitation in generalizing findings to level 2+ transitive dependencies within the npm ecosystem. However, we contend that our sub-analysis, which includes level-1 transitive dependencies, remains a robust approach for assessing generalizability. Level-1 dependencies, being the primary contributors to a package's functionality and characteristics, are pivotal in providing representative insights. Moreover, delving into level 2+ transitive dependencies poses resource challenges due to the extensive nature of dependency networks, making our focus on level-1 dependencies both pragmatic and informative.
Moreover, the sample used for analysis might not be representative of the broader population. If the selection of projects or libraries is biased, the odds ratios calculated may not generalize well to other contexts, limiting the external validity of the findings. Further, the analysis may be sensitive to the timing of data collection. Technological changes, updates, or shifts in development practices over time could impact both technical leverage and vulnerability rates, affecting the stability and generalizability of the odds ratios.
Another potential threat associated with metrics used to measure technical leverage, particularly those reliant on the size of the codebase and its dependencies, lies in their susceptibility to certain challenges. This includes concerns related to issues associated with dependency depth and the static nature of the size metric which may not capture the evolving nature of technical leverage over time.
We acknowledge that our study's conclusions, such as the observed increase in vulnerability risk linked to high technical leverage across various library sizes, may be affected by selection bias in our sample. Our  study was designed to investigate the opportunities and risks associated with high technical leverage, focusing on the top 142 libraries known for their widespread usage and influence. Thus, our findings may not generalize to all libraries in the npm ecosystem. In future research, we plan to include a diverse range of libraries, considering factors such as size, domain, and community support.

\section{Conclusions} \label{conclusion}
To capture the importance of notions introduced by Massacci and Pashchenko \cite{massacci2021technical} on the NPM software ecosystem, we have applied the direct technical leverage and related metrics to 14,042 NPM library releases. Moreover, we extend the study to investigate the impact of including the first level of transitive dependencies on technical leverage.
The results show that small-medium libraries leverage considerably more code from FOSS than large libraries, with developers of small-medium libraries constantly including more dependencies. 
We also find that releases from small-medium libraries with high direct technical leverage tend to ship faster than releases with low direct technical leverage. However, when taking into account the first level of transitive dependencies, the release process is only slightly slower.  It is worth noting that although high direct technical leverage can expedite the release process, it also entails potential risks and costs. 
Libraries that leverage more code from dependencies have, on average, at least 4 times the risk of being affected by vulnerabilities than libraries that leverage less code from dependencies. Moreover, the risk of vulnerabilities increases by 6.7 times when high technical leverage is involved, especially when including the first level of transitive dependencies.

Overall, our study shows that technical leverage is associated with opportunities for shipping more code in less time, but comes at the cost of a higher risk of vulnerabilities. 
Both the opportunities and risks are magnified in NPM, compared to the Maven ecosystem.
Therefore, JavaScript developers should cautiously select their dependencies to avoid staying on the losing end of the trade-off technical leverage entails.   


\section{Data Availability}
We provided a replication package containing all data and scripts used in our study. We refer to the replication package as part of our list of contributions in the introduction. Link: https://zenodo.org/record/6585292

\section{Declarations}
\textbf{Funding and/or Conflicts of interests/Competing interests.} The authors declare that they have no confict of
interest.

\bibliographystyle{abbrv}

\bibliography{bibliography}

\begin{thebibliography}{10}

\bibitem{cohen}
Effect size cohen,
  https://www.statisticssolutions.com/free-resources/directory-of-statistical-analyses/effect-size/,
  Accessed on 01/31/2024.

\bibitem{Libraries}
Libraries-the open source discovery service, https://libraries.io/, Accessed on
  11/09/2022.

\bibitem{snyk}
Snyk website, https://snyk.io/, Accessed on 11/09/2022.

\bibitem{devecosystem}
The state of developer ecosystem,
  https://www.jetbrains.com/lp/devecosystem-2022/, Accessed on 1/25/2024.

\bibitem{snyk_database}
Snyk vulnerability db, https://snyk.io/vuln, Accessed on 1/9/2023.

\bibitem{abdalkareem2017developers}
R.~Abdalkareem, O.~Nourry, S.~Wehaibi, S.~Mujahid, and E.~Shihab.
\newblock Why do developers use trivial packages? an empirical case study on
  npm.
\newblock In {\em Proceedings of the 2017 11th joint meeting on foundations of
  software engineering}, pages 385--395, 2017.

\bibitem{abdalkareem2020impact}
R.~Abdalkareem, V.~Oda, S.~Mujahid, and E.~Shihab.
\newblock On the impact of using trivial packages: An empirical case study on
  npm and pypi.
\newblock {\em Empirical Software Engineering}, 25(2):1168--1204, 2020.

\bibitem{abdellatif_ist2020}
A.~Abdellatif, Y.~Zeng, M.~Elshafei, E.~Shihab, and W.~Shang.
\newblock Simplifying the {Search} of npm {Packages}.
\newblock {\em Information and Software Technology}, 126, 2020.

\bibitem{abdulkareem2021evaluating}
S.~A. Abdulkareem and A.~J. Abboud.
\newblock Evaluating python, c++, javascript and java programming languages
  based on software complexity calculator (halstead metrics).
\newblock In {\em IOP Conference Series: Materials Science and Engineering},
  volume 1076, page 012046. IOP Publishing, 2021.

\bibitem{Alfadel:2021:Python}
M.~Alfadel, D.~E. Costa, and E.~Shihab.
\newblock Empirical analysis of security vulnerabilities in python packages.
\newblock In {\em 2021 IEEE International Conference on Software Analysis,
  Evolution and Reengineering (SANER)}, pages 446--457, 2021.

\bibitem{Alfadel21Dependabot}
M.~Alfadel, D.~E. Costa, E.~Shihab, and M.~Mkhallalati.
\newblock On the use of dependabot security pull requests.
\newblock In {\em 2021 IEEE/ACM 18th International Conference on Mining
  Software Repositories (MSR)}, pages 254--265, 2021.

\bibitem{anderson1962introduction}
T.~W. Anderson.
\newblock An introduction to multivariate statistical analysis.
\newblock Technical report, Wiley New York, 1962.

\bibitem{dataset_link}
Anonymous.
\newblock Opportunities and security risks of technical leverage: A replication
  study on the npm ecosystem | zenodo.
\newblock \url{https://zenodo.org/record/6402982\#.YkctjJPML0o}, Mar 2022.
\newblock (Accessed on 11/31/2022).

\bibitem{azad2019less}
B.~A. Azad, P.~Laperdrix, and N.~Nikiforakis.
\newblock Less is more: quantifying the security benefits of debloating web
  applications.
\newblock In {\em 28th USENIX Security Symposium (USENIX Security 19)}, pages
  1697--1714, 2019.

\bibitem{basili1996reuse}
V.~R. Basili, L.~C. Briand, and W.~L. Melo.
\newblock How reuse influences productivity in object-oriented systems.
\newblock {\em Communications of the ACM}, 39(10):104--116, 1996.

\bibitem{Bogart:21:PracticesPolicies}
C.~Bogart, C.~K\"{a}stner, J.~Herbsleb, and F.~Thung.
\newblock When and how to make breaking changes: Policies and practices in 18
  open source software ecosystems.
\newblock {\em ACM Trans. Softw. Eng. Methodol.}, 30(4), jul 2021.

\bibitem{Paolo:21:Finegrained}
P.~Boldi and G.~Gousios.
\newblock Fine-grained network analysis for modern software ecosystems.
\newblock {\em ACM Trans. Internet Technol.}, 21(1), dec 2020.

\bibitem{chinthanet2021lags}
B.~Chinthanet, R.~G. Kula, S.~McIntosh, T.~Ishio, A.~Ihara, and K.~Matsumoto.
\newblock Lags in the release, adoption, and propagation of npm vulnerability
  fixes.
\newblock {\em Empirical Software Engineering}, 26(3):1--28, 2021.

\bibitem{cogo2021empirical}
F.~R. Cogo, G.~A. Oliva, C.-P. Bezemer, and A.~E. Hassan.
\newblock An empirical study of same-day releases of popular packages in the
  npm ecosystem.
\newblock {\em Empirical Software Engineering}, 26(5):1--42, 2021.

\bibitem{cox2015measuring}
J.~Cox, E.~Bouwers, M.~Van~Eekelen, and J.~Visser.
\newblock Measuring dependency freshness in software systems.
\newblock In {\em 2015 IEEE/ACM 37th IEEE International Conference on Software
  Engineering}, volume~2, pages 109--118. IEEE, 2015.

\bibitem{dashevskyi2018screening}
S.~Dashevskyi, A.~D. Brucker, and F.~Massacci.
\newblock A screening test for disclosed vulnerabilities in foss components.
\newblock {\em IEEE Transactions on Software Engineering}, 45(10):945--966,
  2018.

\bibitem{decan2016topology}
A.~Decan, T.~Mens, and M.~Claes.
\newblock On the topology of package dependency networks: A comparison of three
  programming language ecosystems.
\newblock In {\em Proccedings of the 10th European Conference on Software
  Architecture Workshops}, pages 1--4, 2016.

\bibitem{decan2017empirical}
A.~Decan, T.~Mens, and M.~Claes.
\newblock An empirical comparison of dependency issues in oss packaging
  ecosystems.
\newblock In {\em 2017 IEEE 24th International Conference on Software Analysis,
  Evolution and Reengineering (SANER)}, pages 2--12. IEEE, 2017.

\bibitem{decan2018evolution}
A.~Decan, T.~Mens, and E.~Constantinou.
\newblock On the evolution of technical lag in the npm package dependency
  network.
\newblock In {\em 2018 IEEE International Conference on Software Maintenance
  and Evolution (ICSME)}, pages 404--414. IEEE, 2018.

\bibitem{decan2018impact}
A.~Decan, T.~Mens, and E.~Constantinou.
\newblock On the impact of security vulnerabilities in the npm package
  dependency network.
\newblock In {\em Proceedings of the 15th international conference on mining
  software repositories}, pages 181--191, 2018.

\bibitem{decan2019empirical}
A.~Decan, T.~Mens, and P.~Grosjean.
\newblock An empirical comparison of dependency network evolution in seven
  software packaging ecosystems.
\newblock {\em Empirical Software Engineering}, 24(1):381--416, 2019.

\bibitem{dusing2022analyzing}
J.~D{\"u}sing and B.~Hermann.
\newblock Analyzing the direct and transitive impact of vulnerabilities onto
  different artifact repositories.
\newblock {\em Digital Threats: Research and Practice}, 3(4):1--25, 2022.

\bibitem{fisher1992statistical}
R.~A. Fisher.
\newblock Statistical methods for research workers.
\newblock In {\em Breakthroughs in statistics}, pages 66--70. Springer, 1992.

\bibitem{flauzino2018you}
M.~Flauzino, J.~Ver{\'\i}ssimo, R.~Terra, E.~Cirilo, V.~H. Durelli, and R.~S.
  Durelli.
\newblock Are you still smelling it? a comparative study between java and
  kotlin language.
\newblock In {\em Proceedings of the VII Brazilian symposium on software
  components, architectures, and reuse}, pages 23--32, 2018.

\bibitem{Equifaxd60:online}
J.~Fruhlinger.
\newblock Equifax data breach faq: What happened, who was affected, what was
  the impact? | cso online.
\newblock
  \url{https://www.csoonline.com/article/3444488/equifax-data-breach-faq-what-happened-who-was-affected-what-was-the-impact.html},
  Feb 2020.
\newblock (Accessed on 01/31/2023).

\bibitem{gkortzis2019double}
A.~Gkortzis, D.~Feitosa, and D.~Spinellis.
\newblock A double-edged sword? software reuse and potential security
  vulnerabilities.
\newblock In {\em International Conference on Software and Systems Reuse},
  pages 187--203. Springer, 2019.

\bibitem{gkortzis2021software}
A.~Gkortzis, D.~Feitosa, and D.~Spinellis.
\newblock Software reuse cuts both ways: An empirical analysis of its
  relationship with security vulnerabilities.
\newblock {\em Journal of Systems and Software}, 172:110653, 2021.

\bibitem{grinter1996understanding}
R.~E. Grinter.
\newblock {\em Understanding dependencies: A study of the coordination
  challenges in software development}.
\newblock PhD thesis, University of California, Irvine, 1996.

\bibitem{Joseph_Hejderup}
J.~Hejderup.
\newblock {\em In Dependencies We Trust: How vulnerable are dependencies in
  software modules?}
\newblock PhD thesis, 01 2015.

\bibitem{hejderup2015dependencies}
J.~Hejderup.
\newblock In dependencies we trust: How vulnerable are dependencies in software
  modules?
\newblock 2015.

\bibitem{husain2019codesearchnet}
H.~Husain, H.-H. Wu, T.~Gazit, M.~Allamanis, and M.~Brockschmidt.
\newblock Codesearchnet challenge: Evaluating the state of semantic code
  search.
\newblock {\em arXiv preprint arXiv:1909.09436}, 2019.

\bibitem{imtiaz2021comparative}
N.~Imtiaz, S.~Thorn, and L.~Williams.
\newblock A comparative study of vulnerability reporting by software
  composition analysis tools.
\newblock In {\em Proceedings of the 15th ACM/IEEE International Symposium on
  Empirical Software Engineering and Measurement (ESEM)}, pages 1--11, 2021.

\bibitem{kikas2017structure}
R.~Kikas, G.~Gousios, M.~Dumas, and D.~Pfahl.
\newblock Structure and evolution of package dependency networks.
\newblock In {\em 2017 IEEE/ACM 14th International Conference on Mining
  Software Repositories (MSR)}, pages 102--112. IEEE, 2017.

\bibitem{koishybayev2020mininode}
I.~Koishybayev and A.~Kapravelos.
\newblock Mininode: Reducing the attack surface of node. js applications.
\newblock In {\em 23rd International Symposium on Research in Attacks,
  Intrusions and Defenses (RAID 2020)}, pages 121--134, 2020.

\bibitem{kula2018developers}
R.~G. Kula, D.~M. German, A.~Ouni, T.~Ishio, and K.~Inoue.
\newblock Do developers update their library dependencies?
\newblock {\em Empirical Software Engineering}, 23(1):384--417, 2018.

\bibitem{latendresse2022not}
J.~Latendresse, S.~Mujahid, D.~E. Costa, and E.~Shihab.
\newblock Not all dependencies are equal: An empirical study on production
  dependencies in npm.
\newblock 2023.

\bibitem{lauinger2018thou}
T.~Lauinger, A.~Chaabane, S.~Arshad, W.~Robertson, C.~Wilson, and E.~Kirda.
\newblock Thou shalt not depend on me: Analysing the use of outdated javascript
  libraries on the web.
\newblock {\em arXiv preprint arXiv:1811.00918}, 2018.

\bibitem{lim1994effects}
W.~C. Lim.
\newblock Effects of reuse on quality, productivity, and economics.
\newblock {\em IEEE software}, 11(5):23--30, 1994.

\bibitem{liu2022demystifying}
C.~Liu, S.~Chen, L.~Fan, B.~Chen, Y.~Liu, and X.~Peng.
\newblock Demystifying the vulnerability propagation and its evolution via
  dependency trees in the npm ecosystem.
\newblock In {\em Proceedings of the 44th International Conference on Software
  Engineering}, pages 672--684, 2022.

\bibitem{luszcz2018apache}
J.~Luszcz.
\newblock Apache struts 2: how technical and development gaps caused the
  equifax breach.
\newblock {\em Network Security}, 2018(1):5--8, 2018.

\bibitem{massacci2021technical}
F.~Massacci and I.~Pashchenko.
\newblock Technical leverage in a software ecosystem: Development opportunities
  and security risks.
\newblock In {\em 2021 IEEE/ACM 43rd International Conference on Software
  Engineering (ICSE)}, pages 1386--1397. IEEE, 2021.

\bibitem{mir2023effect}
A.~M. Mir, M.~Keshani, and S.~Proksch.
\newblock On the effect of transitivity and granularity on vulnerability
  propagation in the maven ecosystem.
\newblock {\em arXiv preprint arXiv:2301.07972}, 2023.

\bibitem{mohagheghi2007quality}
P.~Mohagheghi and R.~Conradi.
\newblock Quality, productivity and economic benefits of software reuse: a
  review of industrial studies.
\newblock {\em Empirical Software Engineering}, 12(5):471--516, 2007.

\bibitem{mohagheghi2004empirical}
P.~Mohagheghi, R.~Conradi, O.~M. Killi, and H.~Schwarz.
\newblock An empirical study of software reuse vs. defect-density and
  stability.
\newblock In {\em Proceedings. 26th International Conference on Software
  Engineering}, pages 282--291. IEEE, 2004.

\bibitem{moller2020detecting}
A.~M{\o}ller, B.~B. Nielsen, and M.~T. Torp.
\newblock Detecting locations in javascript programs affected by breaking
  library changes.
\newblock {\em Proceedings of the ACM on Programming Languages},
  4(OOPSLA):1--25, 2020.

\bibitem{nappa2015attack}
A.~Nappa, R.~Johnson, L.~Bilge, J.~Caballero, and T.~Dumitras.
\newblock The attack of the clones: A study of the impact of shared code on
  vulnerability patching.
\newblock In {\em 2015 IEEE symposium on security and privacy}, pages 692--708.
  IEEE, 2015.

\bibitem{andrew_nesbitt_2017_808273}
A.~Nesbitt and B.~Nickolls.
\newblock Libraries.io, open source repository and dependency metadata,
  https://doi.org/10.5281/zenodo.808273, June 2017.

\bibitem{neuburger2021trends}
J.~D. Neuburger, J.~P. Mollod, and L.~Proskauer~Rose.
\newblock Trends in privacy and data security: 2020.
\newblock {\em Proskauer}, 2021.

\bibitem{nielsen2021semantic}
B.~B. Nielsen, M.~T. Torp, and A.~M{\o}ller.
\newblock Semantic patches for adaptation of javascript programs to evolving
  libraries.
\newblock In {\em 2021 IEEE/ACM 43rd International Conference on Software
  Engineering (ICSE)}, pages 74--85. IEEE, 2021.

\bibitem{pashakhanloo2022pacjam}
P.~Pashakhanloo, A.~Machiry, H.~Choi, A.~Canino, K.~Heo, I.~Lee, and M.~Naik.
\newblock Pacjam: Securing dependencies continuously via package-oriented
  debloating.
\newblock In {\em Proceedings of the 2022 ACM on Asia Conference on Computer
  and Communications Security}, pages 903--916, 2022.

\bibitem{pashchenko2018vulnerable}
I.~Pashchenko, H.~Plate, S.~E. Ponta, A.~Sabetta, and F.~Massacci.
\newblock Vulnerable open source dependencies: Counting those that matter.
\newblock In {\em Proceedings of the 12th ACM/IEEE International Symposium on
  Empirical Software Engineering and Measurement}, pages 1--10, 2018.

\bibitem{pashchenko2020qualitative}
I.~Pashchenko, D.-L. Vu, and F.~Massacci.
\newblock A qualitative study of dependency management and its security
  implications.
\newblock In {\em Proceedings of the 2020 ACM SIGSAC Conference on Computer and
  Communications Security}, pages 1513--1531, 2020.

\bibitem{pham2010detection}
N.~H. Pham, T.~T. Nguyen, H.~A. Nguyen, and T.~N. Nguyen.
\newblock Detection of recurring software vulnerabilities.
\newblock In {\em Proceedings of the IEEE/ACM international conference on
  Automated software engineering}, pages 447--456, 2010.

\bibitem{Ponta:21:Bloated}
S.~E. Ponta, W.~Fischer, H.~Plate, and A.~Sabetta.
\newblock The used, the bloated, and the vulnerable: Reducing the attack
  surface of an industrial application.
\newblock In {\em 2021 IEEE International Conference on Software Maintenance
  and Evolution (ICSME)}, pages 555--558, 2021.

\bibitem{prana2021out}
G.~A.~A. Prana, A.~Sharma, L.~K. Shar, D.~Foo, A.~E. Santosa, A.~Sharma, and
  D.~Lo.
\newblock Out of sight, out of mind? how vulnerable dependencies affect
  open-source projects.
\newblock {\em Empirical Software Engineering}, 26(4):1--34, 2021.

\bibitem{quach2019bloat}
A.~Quach and A.~Prakash.
\newblock Bloat factors and binary specialization.
\newblock In {\em Proceedings of the 3rd ACM Workshop on Forming an Ecosystem
  Around Software Transformation}, pages 31--38, 2019.

\bibitem{raemaekers2014semantic}
S.~Raemaekers, A.~Van~Deursen, and J.~Visser.
\newblock Semantic versioning versus breaking changes: A study of the maven
  repository.
\newblock In {\em 2014 IEEE 14th International Working Conference on Source
  Code Analysis and Manipulation}, pages 215--224. IEEE, 2014.

\bibitem{sawant2018understanding}
A.~A. Sawant, M.~Aniche, A.~van Deursen, and A.~Bacchelli.
\newblock Understanding developers' needs on deprecation as a language feature.
\newblock In {\em Proceedings of the 40th International Conference on Software
  Engineering}, pages 561--571, 2018.

\bibitem{2020Soft27:online}
Sonatype.
\newblock 2020 software supply chain report.
\newblock
  \url{https://www.sonatype.com/resources/white-paper-state-of-the-software-supply-chain-2020},
  August 2020.
\newblock (Accessed on 01/31/2023).

\bibitem{soto2021longitudinal}
C.~Soto-Valero, T.~Durieux, and B.~Baudry.
\newblock A longitudinal analysis of bloated java dependencies.
\newblock In {\em Proceedings of the 29th ACM Joint Meeting on European
  Software Engineering Conference and Symposium on the Foundations of Software
  Engineering}, pages 1021--1031, 2021.

\bibitem{soto2022coverage}
C.~Soto-Valero, T.~Durieux, N.~Harrand, and B.~Baudry.
\newblock Coverage-based debloating for java bytecode.
\newblock {\em ACM Computing Surveys (CSUR)}, 2022.

\bibitem{soto2021comprehensive}
C.~Soto-Valero, N.~Harrand, M.~Monperrus, and B.~Baudry.
\newblock A comprehensive study of bloated dependencies in the maven ecosystem.
\newblock {\em Empirical Software Engineering}, 26(3):1--44, 2021.

\bibitem{szumilas2010explaining}
M.~Szumilas.
\newblock Explaining odds ratios.
\newblock {\em Journal of the Canadian academy of child and adolescent
  psychiatry}, 19(3):227, 2010.

\bibitem{turcotte2022stubbifier}
A.~Turcotte, E.~Arteca, A.~Mishra, S.~Alimadadi, and F.~Tip.
\newblock Stubbifier: debloating dynamic server-side javascript applications.
\newblock {\em Empirical Software Engineering}, 27(7):161, 2022.

\bibitem{yu2022pbdiff}
L.~Yu, Y.~Lu, Y.~Shen, J.~Zhao, and J.~Zhao.
\newblock Pbdiff: Neural network based program-wide diffing method for
  binaries.
\newblock {\em Mathematical Biosciences and Engineering}, 19(3):2774--2799,
  2022.

\bibitem{Stan:2022:OpenSource}
S.~Zajdel, D.~E. Costa, and H.~Mili.
\newblock Open source software: An approach to controlling usage and risk in
  application ecosystems, 2022.

\bibitem{zampetti2019study}
F.~Zampetti, G.~Bavota, G.~Canfora, and M.~Di~Penta.
\newblock A study on the interplay between pull request review and continuous
  integration builds.
\newblock In {\em 2019 IEEE 26th International Conference on Software Analysis,
  Evolution and Reengineering (SANER)}, pages 38--48. IEEE, 2019.

\bibitem{zapata2018towards}
R.~E. Zapata, R.~G. Kula, B.~Chinthanet, T.~Ishio, K.~Matsumoto, and A.~Ihara.
\newblock Towards smoother library migrations: A look at vulnerable dependency
  migrations at function level for npm javascript packages.
\newblock In {\em 2018 IEEE International Conference on Software Maintenance
  and Evolution (ICSME)}, pages 559--563. IEEE, 2018.

\bibitem{zar2005spearman}
J.~H. Zar.
\newblock Spearman rank correlation.
\newblock {\em Encyclopedia of biostatistics}, 7, 2005.

\bibitem{zerouali2018empirical}
A.~Zerouali, E.~Constantinou, T.~Mens, G.~Robles, and J.~Gonz{\'a}lez-Barahona.
\newblock An empirical analysis of technical lag in npm package dependencies.
\newblock In {\em International Conference on Software Reuse}, pages 95--110.
  Springer, 2018.

\bibitem{zerouali2019impact}
A.~Zerouali, V.~Cosentino, T.~Mens, G.~Robles, and J.~M. Gonzalez-Barahona.
\newblock On the impact of outdated and vulnerable javascript packages in
  docker images.
\newblock In {\em 2019 IEEE 26th International Conference on Software Analysis,
  Evolution and Reengineering (SANER)}, pages 619--623. IEEE, 2019.

\bibitem{zerouali2021impact}
A.~Zerouali, T.~Mens, A.~Decan, and C.~De~Roover.
\newblock On the impact of security vulnerabilities in the npm and rubygems
  dependency networks.
\newblock {\em arXiv preprint arXiv:2106.06747}, 2021.

\bibitem{zerouali2019diversity}
A.~Zerouali, T.~Mens, G.~Robles, and J.~M. Gonzalez-Barahona.
\newblock On the diversity of software package popularity metrics: An empirical
  study of npm.
\newblock In {\em 2019 IEEE 26th international conference on software analysis,
  Evolution and Reengineering (SANER)}, pages 589--593. IEEE, 2019.

\bibitem{zimmermann2019small}
M.~Zimmermann, C.-A. Staicu, C.~Tenny, and M.~Pradel.
\newblock Small world with high risks: A study of security threats in the npm
  ecosystem.
\newblock In {\em 28th $\{$USENIX$\}$ Security Symposium ($\{$USENIX$\}$
  Security 19)}, pages 995--1010, 2019.

\end{thebibliography}

\end{document}